\documentclass[twocolumn]{aastex631}

\usepackage{amsmath}
\usepackage{ulem}

\shorttitle{JWST and ALMA study of a starburst galaxy at $z=1.45$}
\shortauthors{Liu et al.}

\graphicspath{{./}{figures/}}

\begin{document}

\title{JWST and ALMA discern the assembly of structural and obscured components in a high-redshift starburst galaxy}

\correspondingauthor{Zhaoxuan Liu}
\email{zhaoxuan.liu@ipmu.jp}

\author[0000-0002-9252-114X]{Zhaoxuan Liu}
\affiliation{Kavli Institute for the Physics and Mathematics of the Universe (Kavli IPMU, WPI), UTIAS, Tokyo Institutes for Advanced Study, University of Tokyo, Chiba, 277-8583, Japan}
\affiliation{Department of Astronomy, School of Science, The University of Tokyo, 7-3-1 Hongo, Bunkyo, Tokyo 113-0033, Japan}
\affiliation{Center for Data-Driven Discovery, Kavli IPMU (WPI), UTIAS, The University of Tokyo, Kashiwa, Chiba 277-8583, Japan}

\author[0000-0002-0000-6977]{John D. Silverman}
\affiliation{Kavli Institute for the Physics and Mathematics of the Universe (Kavli IPMU, WPI), UTIAS, Tokyo Institutes for Advanced Study, University of Tokyo, Chiba, 277-8583, Japan}
\affiliation{Department of Astronomy, School of Science, The University of Tokyo, 7-3-1 Hongo, Bunkyo, Tokyo 113-0033, Japan}
\affiliation{Center for Data-Driven Discovery, Kavli IPMU (WPI), UTIAS, The University of Tokyo, Kashiwa, Chiba 277-8583, Japan}
\affiliation{Center for Astrophysical Sciences, Department of Physics \& Astronomy, Johns Hopkins University, Baltimore, MD 21218, USA}

\author[0000-0002-3331-9590]{Emanuele Daddi}
\affiliation{Université Paris-Saclay, Université Paris Cité, CEA, CNRS, AIM, F-91191 Gif-sur-Yvette, France}

\author[0000-0001-9369-1805]{Annagrazia Puglisi}
\affiliation{School of Physics and Astronomy, University of Southampton, Highfield SO17 1BJ, UK}

\author[0000-0002-7093-7355]{Alvio Renzini}
\affiliation{INAF - Osservatorio Astronomico di Padova, Vicolo dell'Osservatorio 5, I-35122 Padova, Italy}

\author[0000-0001-9215-7053]{Boris S. Kalita}
\affiliation{Kavli Institute for the Physics and Mathematics of the Universe (Kavli IPMU, WPI), UTIAS, Tokyo Institutes for Advanced Study, University of Tokyo, Chiba, 277-8583, Japan}
\affiliation{Center for Data-Driven Discovery, Kavli IPMU (WPI), UTIAS, The University of Tokyo, Kashiwa, Chiba 277-8583, Japan}
\affiliation{Kavli Institute for Astronomy and Astrophysics, Peking University, Beijing 100871, P. R. China}

\author[0000-0001-9187-3605]{Jeyhan S. Kartaltepe}
\affiliation{Laboratory for Multiwavelength Astrophysics, School of Physics and Astronomy, Rochester Institute of Technology, 84 Lomb Memorial Drive, Rochester, NY 14623, USA}

\author[0000-0001-9044-1747]{Daichi Kashino}
\affiliation{National Astronomical Observatory of Japan, 2-21-1 Osawa, Mitaka, Tokyo 181-8588, Japan}

\author[0000-0002-9415-2296]{Giulia Rodighiero}
\affiliation{INAF - Osservatorio Astronomico di Padova, Vicolo dell'Osservatorio 5, I-35122 Padova, Italy}
\affiliation{Department of Physics and Astronomy, Università degli Studi di Padova, Vicolo dell'Osservatorio 3, I-35122, Padova, Italy}

\author[0000-0002-0303-499X]{Wiphu Rujopakarn}
\affiliation{Department of Physics, Faculty of Science, Chulalongkorn University, 254 Phayathai Road, Pathumwan, Bangkok 10330, Thailand}
\affiliation{National Astronomical Research Institute of Thailand, 260 Moo 4, T. Donkaew, A. Maerim, Chiangmai 50180, Thailand}

\author[0000-0002-3560-1346]{Tomoko L. Suzuki}
\affiliation{Kavli Institute for the Physics and Mathematics of the Universe (Kavli IPMU, WPI), UTIAS, Tokyo Institutes for Advanced Study, University of Tokyo, Chiba, 277-8583, Japan}
\affiliation{Center for Data-Driven Discovery, Kavli IPMU (WPI), UTIAS, The University of Tokyo, Kashiwa, Chiba 277-8583, Japan}

\author[0009-0003-4742-7060]{Takumi S. Tanaka}
\affiliation{Kavli Institute for the Physics and Mathematics of the Universe (Kavli IPMU, WPI), UTIAS, Tokyo Institutes for Advanced Study, University of Tokyo, Chiba, 277-8583, Japan}
\affiliation{Department of Astronomy, School of Science, The University of Tokyo, 7-3-1 Hongo, Bunkyo, Tokyo 113-0033, Japan}
\affiliation{Center for Data-Driven Discovery, Kavli IPMU (WPI), UTIAS, The University of Tokyo, Kashiwa, Chiba 277-8583, Japan}

\author[0000-0001-6477-4011]{Francesco Valentino}
\affiliation{Cosmic Dawn Center (DAWN), Denmark}
\affiliation{Niels Bohr Institute, University of Copenhagen, Jagtvej 128, DK-2200 Copenhagen N, Denmark}

\author[0000-0001-6102-9526]{Irham Taufik Andika}
\affiliation{Physik-Department, Technische Universität München, James-Franck-Str. 1, D-85748 Garching, Germany}

\author[0000-0002-0930-6466]{Caitlin M. Casey}
\affiliation{The University of Texas at Austin, 2515 Speedway Boulevard Stop C1400, Austin, TX 78712, USA}
\affiliation{Cosmic Dawn Center (DAWN), Denmark}

\author[0000-0002-9382-9832]{Andreas Faisst}
\affiliation{Caltech/IPAC, 1200 E. California Blvd. Pasadena, CA 91125, USA}

\author[0000-0002-3560-8599]{Maximilien Franco}
\affiliation{The University of Texas at Austin, 2515 Speedway Boulevard Stop C1400, Austin, TX 78712, USA}

\author[0000-0002-0236-919X]{Ghassem Gozaliasl}
\affiliation{Department of Computer Science, Aalto University, PO Box 15400, Espoo, FI-00 076, Finland}
\affiliation{Department of Physics, Faculty of Science, University of Helsinki, 00014-Helsinki, Finland}

\author[0000-0001-9885-4589]{Steven Gillman}
\affiliation{Cosmic Dawn Center (DAWN), Denmark}
\affiliation{DTU Space, Technical University of Denmark, Elektrovej, Building 328, 2800, Kgs. Lyngby, Denmark}

\author[0000-0003-4073-3236]{Christopher C. Hayward}
\affiliation{Center for Computational Astrophysics, Flatiron Institute, 162 Fifth Avenue, New York, NY 10010, USA}

\author[0000-0002-6610-2048]{Anton M. Koekemoer}
\affiliation{Space Telescope Science Institute, 3700 San Martin Drive, Baltimore, MD 21218, USA}

\author[0000-0002-5588-9156]{Vasily Kokorev}
\affiliation{Kapteyn Astronomical Institute, University of Groningen, P.O. Box 800, 9700AV Groningen, The Netherlands}

\author[0000-0003-3216-7190]{Erini Lambrides}
\affiliation{NASA Goddard Space Flight Center, Code 662, Greenbelt, MD 20771, USA}

\author[0000-0002-2419-3068]{Minju M. Lee}
\affiliation{Cosmic Dawn Center (DAWN), Denmark}
\affiliation{DTU Space, Technical University of Denmark, Elektrovej, Building 328, 2800, Kgs. Lyngby, Denmark}

\author[0000-0002-4872-2294]{Georgios E. Magdis}
\affiliation{Cosmic Dawn Center (DAWN), Denmark}
\affiliation{DTU-Space, Technical University of Denmark, Elektrovej 327, DK2800 Kgs. Lyngby, Denmark}
\affiliation{Niels Bohr Institute, University of Copenhagen, Jagtvej 128, DK-2200 Copenhagen N, Denmark}

\author[0000-0003-0129-2079]{Santosh Harish}
\affiliation{Laboratory for Multiwavelength Astrophysics, School of Physics and Astronomy, Rochester Institute of Technology, 84 Lomb Memorial Drive, Rochester, NY 14623, USA}

\author[0000-0002-9489-7765]{Henry Joy McCracken}
\affiliation{Institut d'Astrophysique de Paris, UMR 7095, CNRS, and Sorbonne Université, 98 bis boulevard Arago, F-75014 Paris, France}

\author[0000-0002-4485-8549]{Jason Rhodes}
\affiliation{Jet Propulsion Laboratory, California Institute of Technology, 4800 Oak Grove Drive, Pasadena, CA 91001, USA}

\author[0000-0002-7087-0701]{Marko Shuntov}
\affiliation{Institut d'Astrophysique de Paris, UMR 7095, CNRS, and Sorbonne Université, 98 bis boulevard Arago, F-75014 Paris, France}

\author[0000-0001-8917-2148]{ Xuheng Ding}
\affiliation{Kavli Institute for the Physics and Mathematics of the Universe (Kavli IPMU, WPI), UTIAS, Tokyo Institutes for Advanced Study, University of Tokyo, Chiba, 277-8583, Japan}
\affiliation{Center for Data-Driven Discovery, Kavli IPMU (WPI), UTIAS, The University of Tokyo, Kashiwa, Chiba 277-8583, Japan}

\begin{abstract}

We present observations and analysis of the starburst, PACS-819, at \hbox{$z=1.45$} ($M_*=10^{10.7}$ M$_{
\odot}$), using high-resolution ($0^{\prime \prime}.1$; 0.8 kpc) ALMA and multi-wavelength JWST images from the COSMOS-Web program. Dissimilar to HST/ACS images in the rest-frame UV, the redder NIRCam and MIRI images reveal a smooth central mass concentration and spiral-like features, atypical for such an intense starburst. Through dynamical modeling of the CO (J=5--4) emission with ALMA, PACS-819 is rotation-dominated thus consistent with a disk-like nature. However, kinematic anomalies in CO and asymmetric features in the bluer JWST bands (e.g., F150W) support a more disturbed nature likely due to interactions. The JWST imaging further enables us to map the distribution of stellar mass and dust attenuation, thus clarifying the relationships between different structural components, not discernable in the previous HST images. The CO (J=5--4) and FIR dust continuum emission are co-spatial with a heavily-obscured starbursting core ($<1$ kpc) which is partially surrounded by much less obscured star-forming structures including a prominent arc, possibly a tidally-distorted dwarf galaxy, and a massive clump (detected in CO), likely a recently accreted low-mass satellite. With spatially-resolved maps, we find a high molecular gas fraction in the central area reaching $\sim3$ ($M_{\text{gas}}$/$M_*$) and short depletion times ($M_{\text{gas}}/SFR\sim$ 120 Myrs) across the entire system. These observations provide insights into the complex nature of starbursts in the distant Universe and underscore the wealth of complementary information from high-resolution observations with both ALMA and JWST.

\end{abstract}

\keywords{Star formation(1569) --- Starburst galaxies(1570) --- High-redshift galaxies(734) --- Interstellar medium(847)}

\section{Introduction} \label{sec:intro}

What triggers a starburst galaxy? Outliers (i.e., starburst galaxies) are known to exist that are highly elevated above the typical star-forming population. The correlation between stellar mass ($M_{*}$) and star formation rate (SFR) defines the main sequence (MS) of star-forming galaxies which is our reference for identifying these outliers. Even though their contribution to the global SFR density is believed to be subdominant, even in the distant Universe ($\sim$10\%; \citealt{rodighiero_lesser_2011}), starbursts still deserve investigation of their physical properties and triggering mechanisms since they represent a crucial phase in the evolution of massive galaxies \citep[e.g.,][]{Lotz2008,lackner_late-stage_2014}. 

In the local Universe, these starbursts are the ultra-luminous infrared galaxies (ULIRGs) triggered by major-mergers \citep{solomon_molecular_1997, sanders_ultraluminous_1988}. They typical exhibit high specific star formation rates (sSFR = SFR/$M_{*}$), compact molecular gas reservoirs and short gas depletion times ($\tau_{\mathrm{depl}}$ =  $M_{\mathrm{gas}}/{\mathrm{SFR}}$). 

However, in the distant Universe, whether the physical condition of a starburst is consistent with $z\sim0$ is still being investigated. According to simulations and related theoretical analysis  \citep{hernquist_tidal_1989, mihos_ultraluminous_1994, barnes_transformations_1996, dekel_cold_2009, moreno_spatially_2021}, the distant starbursts could be triggered by mergers and also by (violent) disk instabilities ultimately due to the high gas fraction in galaxies in the early Universe \citep[e.g.,][]{tacconi_evolution_2020}. In a focused investigation of high-redshift starbursts with ALMA \citep{silverman_higher_2015,silverman_molecular_2018,silverman_concurrent_2018}, a higher efficiency of converting gas to stars is found at $z\sim$1.6 \citep[see also][]{tacconi_phibss_2018,tacconi_evolution_2020}. On the other hand, the elevated SFRs of distant starbursts may be due to their enhanced gas fractions, not solely driven by a higher efficiency \citep{scoville_ism_2016,scoville_cosmic_2023}. 

In addition, some starbursts remain hidden within the Main Sequence (MS), displaying short depletion times \citep{magdis_evolving_2012, elbaz_starbursts_2018, gomez-guijarro_goods-alma_2022} and compact cores in the submillimeter \citep{puglisi_main_2019, puglisi_sub-millimetre_2021}. These studies hint that a resolved study associating the stellar and gas is needed. However, a challenge in earlier studies was combining the properties of the ISM with those of stars at spatially resolved scales. This was largely due to the significant obscuration caused by dust, making it difficult to derive spatially resolved attributes from instruments like HST and ground-based telescopes. 

With the advent of JWST, it is possible to shed light on the stellar populations hidden behind the dust veil in the distant Universe. Recent JWST research \citep{2023arXiv230707599L} indicates that starbursts within the MS feature a compact starburst core and a surrounding massive disk with a reduced sSFR. Given these advancements, it's imperative to reexamine off-MS starbursts using more in-depth, high-resolution multi-wavelength observations to understand better the mechanisms driving their starburst activity and departure from the MS.

In this study, we present an extreme starburst event (11$\times$ above the MS), PACS-819, utilizing high-resolution observations with JWST/NIRCam+MIRI images from COSMOS-Web \citep{casey_cosmos-web_2023} and ALMA Band 6. With a spatial resolution of approximately $0.1^{\prime\prime}$ and high signal-to-noise emission with ALMA, we successfully resolve the carbon monoxide $^{12}$CO transition $5\rightarrow4$ and dust continuum emission, enabling us to delve into the intricate morphology and kinematics of its compact center. Additionally, we identify a clump through its presence in CO emission and its impact on disk kinematics. The JWST observations enable us to produce stellar mass and extinction maps which are important in unveiling the nature of PACS-819, particularly the heavily obscured central region. By investigating this galaxy in such detail, we aim to advance our understanding of the trigger of the distant starbursts. Throughout this work, we follow the standard $\Lambda\mathrm{CDM}$ model by assuming $H_{0}=70 \mathrm{~km} \mathrm{~s}^{-1} \mathrm{Mpc}^{-1}, \Omega_{\Lambda}=0.7, \Omega_{\mathrm{M}}=0.3$. We use a \citet{chabrier_galactic_2003} IMF for estimating the stellar masses and SFRs.

\section{Target, data and analysis}

\subsection{PACS-819: a starburst galaxy at $z=1.45$} \label{sec:Sample}

\begin{figure*}[h]
\centering
\includegraphics[width=16cm]{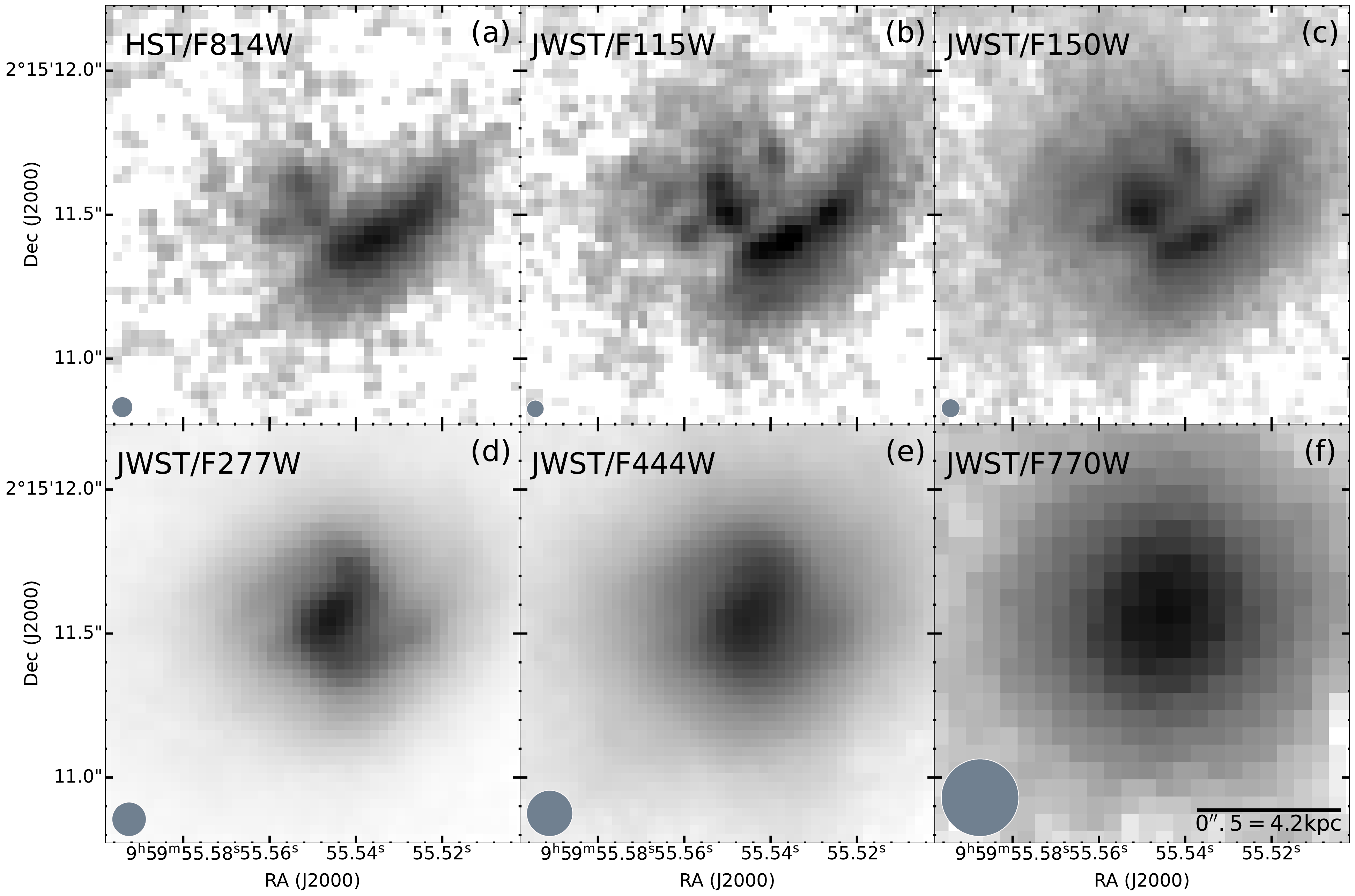}
\caption{HST and JWST observations of PACS-819. Panels (a--f) are ordered from short to long wavelength including HST ACS/F814W and five JWST (NIRCam: F115W, F150W, F277W, and F444W; MIRI: F770W) bands. In each panel, the north is up and east is to the left. All six panels show the same region of the sky and 1$^{\prime\prime}$ = 8.45 kpc in projected distance. The small circle represents the spatial resolution.} 
\label{fig:obs819}
\end{figure*}

PACS-819 is a distant starburst galaxy at $z$ = 1.445, detected by {\it Herschel}/PACS \citep{rodighiero_lesser_2011} and included in our FMOS-COSMOS near-infrared spectroscopic survey of star-forming galaxies with the Subaru Telescope \citep{Kashino2019}. Its stellar mass $M_*$ is estimated to be $10^{10.7 \pm 0.1}~M_{\odot}$ from a comprehensive SED fitting conducted by \citet{liu_co_2021}. The SFR is estimated to be 533$^{+68}_{-60}$  M$_{\odot}$ yr$^{-1}$, derived from $L_{\text{TIR}}$ \citep{silverman_molecular_2018}. These measurements place PACS-819 just over 10 times above the MS \citep{speagle_highly_2014}, given t = 4.32 Gyr (z = 1.45). Its starburst-like qualities (i.e., high gas fraction and star-forming efficiency) are based on CO (J=2--1) detections with S/N $>$ 5 using ALMA \citep{silverman_fmos-cosmos_2015}. Notably, PACS-819 exhibits signs of a merger in HST F814W, with multiple UV/optical emission regions as depicted in Figure~\ref{fig:obs819}a. Along with these properties, the high S/N CO (J=2--1) detection prompted our request for a higher-resolution observation with ALMA.

\subsection{JWST (+HST) imaging from COSMOS-Web}

PACS-819 was observed with JWST/NIRCam on April 9th, 2023 as part of the large COSMOS-Web program (PIs: Casey \& Kartaltepe; \citealt{casey_cosmos-web_2023}; GO \#1727) of JWST Cycle 1. Images with four NIRCam filters (F115W, F150W, F277W, and F444W) are available. Additionally, on April 10th, PACS-819 was observed with MIRI/F770W. The 5$\sigma$ depths of these observations are 27.13, 27.35, 27.99, 27.83, and 25.70 magnitudes, respectively. The data used in this study is recalibrated by the COSMOS-Web team and reduced to a spatial scale of 30 mas/pixel (Franco in prep.). 

To supplement the rest-frame optical filters, we also include previous observations conducted with HST/ACS. PACS-819 was observed in F606W and F814W as a part of the HST Cycle 19 (PI: A. Riess; Program \#12461). It was also covered by the COSMOS project \citep{scoville_cosmos_2007} in F814W \citep{koekemoer_cosmos_2007}.  We utilized the F606W image at 40 mas/pixel, as reduced using the \textsc{grizli} software \citep{2023zndo...7712834B}, and the F814W image at 30 mas/pixel by the COSMOS-Web team following the approaches first described in \citet{koekemoer_candels_2011}. Together, this multi-wavelength data set from HST and JWST provides a comprehensive view, crucial for dissecting the complex morphology of this dusty starburst galaxy, particularly in the rest-frame UV and optical.

\subsubsection{SED fitting}
\label{sec:sed}
To address the source complexity and make full use of the high-resolution images, we employ spatially resolved spectral energy distribution (SED) fitting on a pixel-by-pixel basis using two different software packages: \texttt{pixedfit} \citep{abdurrouf_introducing_2021} and \texttt{bagpipes} \citep{carnall_inferring_2018}. The tool \texttt{pixedfit} provides us with a workflow for pixel-to-pixel SED fitting which consists of several individual modules. We utilize \texttt{piXedfit\_images} to handle the image processing, downgrading all images to match the PSF (point spread function) of F444W and the pixel scale of F606W at 40 mas/pixel. For convolving the images, HST PSFs were generated by averaging the profiles of field stars. The JWST PSFs were generated using PSFEx \citep{2011ASPC..442..435B}. The re-sampled images physically probe the scale of a single resolved component with 4 pixels at 40 mas/pixel resolution. After correcting for galactic extinction, the photometric data cube, including the use of a segmentation map generated by the image module, was passed to the binning module to increase the signal-to-noise of some spatial elements. \texttt{piXedfit\_bin} \citep{abdurrouf_introducing_2021} bins pixels with similar SED shapes to meet the S/N requirements and can smooth the data. When binning, we requested all bins to have an S/N $\ge$ 10 in F150W, F277W and F444W while the smallest bin size $D_{min,bin}$ is 1 pixel, i.e., no binning, to not sacrifice the number of pixels in the central bright regions for comparison with the ALMA images having a similar resolution. We did not include the dust continuum because it is hard to derive an accurate FIR model with only one band.

We then passed the rebinned photometric cubes to \texttt{bagpipes} for SED fitting, implementing a simple model due to the limited number of bands. We introduced 10\% systematic uncertainty, as suggested by \citet{abdurrouf_understanding_2017}. Since the metallicity has limited impact on our primary objective to assess the stellar mass \citep{osborne_strategies_2024}, we fix the metallicity to solar (Z=0.02) to simplify our model when only six bands are available. We first applied a delayed-exponential star-formation history while allowing $\tau$ to vary. Dust modeling follows the method of \citet{calzetti_dust_2000} with $A_V$ varying between 0 and 4 magnitudes since the range in attenuation is likely large as suggested by the color map of PACS-819 (Figure~\ref{fig:rgb}). Additionally, we fixed the ionization parameter U for the nebular emission to be $10^{-3}$ as the default. We constructed the UVJ color map with the best-fit SEDs of each bin to confirm that the entire galaxy is star-forming with the classification scheme of \citet{whitaker_star_2012}. Then, we further simplified the model to a constant SFH, with a variable turning-on time from 100 Myr to the age of the Universe and a turning-off time at 1 Myr, keeping other priors the same.

\subsection{ALMA observations}
\label{sec:Obs}

With our ALMA Cycle 4 program (Proposal \#2016.1.01426.S; PI J. Silverman), we acquired high- and low-resolution ALMA observations of three starbursts well above the MS: PACS-819, PACS-830, and PACS-787 \citep{silverman_concurrent_2018}. Here, we focus on PACS-819 which was observed with 1.46 hrs of total on-source integration time with 45/40 12 m antennas thus resulting in high-resolution observations having a beam size of $0^{\prime \prime}.12 \times 0^{\prime \prime}.09$ in Band 6 to resolve the CO (J=5--4) emission line ($\nu_{\mathrm{rest}}$ = 576.27 GHz) and the underlying continuum ($\lambda_{\rm obs}$ = 1.3 mm). In the low-resolution configuration, PACS-819 was observed with 22.2 minutes with 42/43 12 m antennas to capture the total flux of the CO (J=5--4) and underlying continuum. The resulting beam size is $0^{\prime \prime}.83 \times 0^{\prime \prime}.68$. The summary of the ALMA observations is provided in Table~\ref{tab:observations}.

We use the Common Astronomy Software Applications (CASA 4.7.2) to carry out the reduction and recalibration of the raw data. Our measurements and analysis are processed by CASA 6.2.1 and python package \texttt{spectral\_cube} as described in the following subsections.

\subsubsection{CO (J=5--4)} \label{subsec:54e}

When constructing the high-resolution image cube, we applied a Briggs \citep{1995AAS...18711202B} weighting factor of 0.5 to achieve a higher resolution and optimize the sensitivity for substructures in the gas distribution. The cube was subsequently rebinned to 20 $\text{km}\,\text{s}^{-1}$ (4 times the native resolution) to enhance the signal-to-noise ratio. We then collapse our high-resolution CO (J=5--4) detection over its full spectral range by manually setting $\Delta v$ in velocity, as listed in Table \ref{tab:properties2}. For the low-resolution cube, we rebinned it to 35 $\text{km}\,\text{s}^{-1}$ (7 times the native resolution) to detect internal structures.

To perform flux measurements, we used the \texttt{imfit} task to generate a model based on an elliptical Gaussian function. A point-source component is added to model a potential unresolved central component; however, only 5\% of the emission is associated with this component at the 3$\sigma$ significance level. The fit results, excluding the point source component, are listed in Table \ref{tab:properties2}.

\subsubsection{Continuum} \label{subsec:con}
To identify and exclude emission lines in the continuum analysis, we initially flagged the frequency range of 4$\times$FWHM (Full Width at Half Maximum) centered at the line centroid in the dirty cubes, based on the ALMA data release products. Subsequently, we employed the \texttt{tclean} algorithm in multifrequency synthesis (\texttt{mfs}, \citealt{1990MNRAS.246..490C}) mode, covering all channels without emission lines. We also used \texttt{multiscale} deconvolver to detect the extended emission. For high-resolution images, we utilized the \texttt{auto-multithresh} \citep{cornwell_multiscale_2008} mask and opted for natural weighting to maximize the signal-to-noise ratio and a 0.5 Briggs weighting for a higher resolution. The resulting restored beams are 0$^{\prime \prime}$.16 $\times$ 0$^{\prime \prime}$.11 and 0$^{\prime \prime}$.13 $\times$ 0$^{\prime \prime}$.09 respectively. For the low-resolution observation, we only applied the 0.5 Briggs weighting for its high signal-to-noise ratio. The restored beam size is 0$^{\prime \prime}$.83 $\times$ 0$^{\prime \prime}$.69.
 
To derive morphological information and far-infrared flux, we utilized the \texttt{imfit} task in CASA, employing multiple 2D Gaussian models with the natural weighted image. A Gaussian model combined with a point source yielded a satisfactory result to describe its compact core.

\subsection{AGN}
\citet{huang_diverse_2023} demonstrate that some distant ULIRGs at $z\sim2$ have an AGN contribution. Therefore, we searched for evidence supporting an association of the central starburst with AGN activity in PACS-819. There is no detection in the previous X-ray survey with $Chandra$ and XMM-$Newton$ in the COSMOS field \citep{cappelluti_xmm-newton_2009,civano_chandra_2016,marchesi_chandra_2016}. In addition, the SED fitting using mid-infrared AGN SEDs from \citep{mullaney_defining_2011} with global multi-wavelength photometry by \citet{liu_automated_2019} shows no significant AGN contribution. Even though this starburst lies on the star-forming side near the boundary between SFGs and AGNs in the BPT diagram, analyzed in \citet{silverman_higher_2015} with FMOS spectra, we conclude that the AGN activity is not likely to be present.

\begin{figure*}
\epsscale{1.2}
\plotone{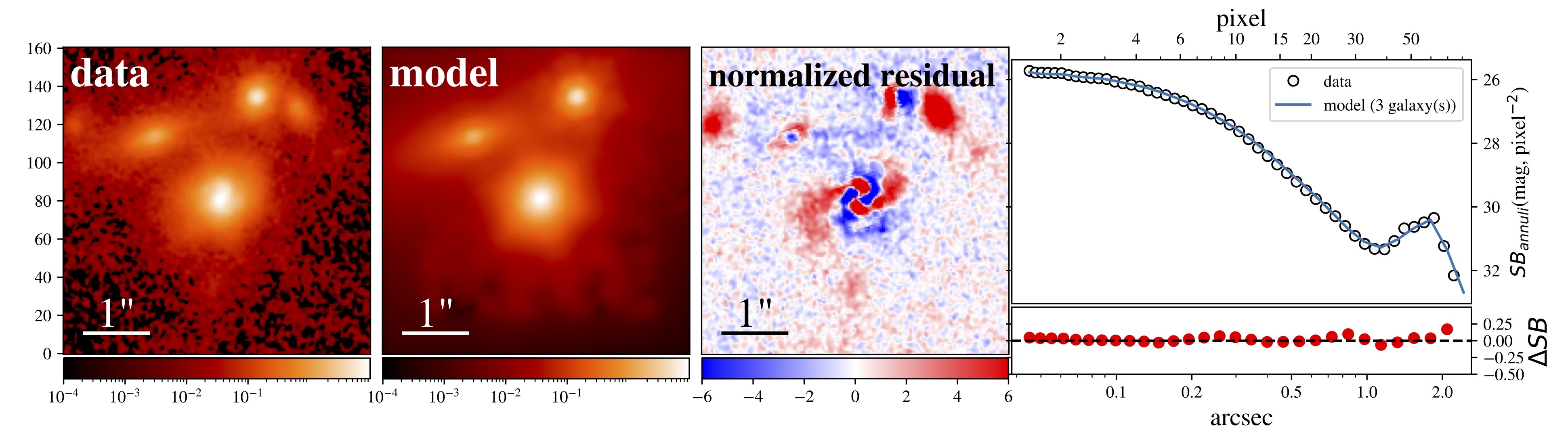}
\caption{S\'ersic model fit to the 2D emission in F444W using Galight with panels as follows from left to right: science image, model image, residual image (data--model/$\sigma$) and 1D surface brightness profile. The residual image shows spiral features typical for a star-forming galaxy not undergoing a major merger.}
\label{fig:img_model_fit}
\end{figure*}

\begin{figure}[h]
\centering
\includegraphics[width=8cm]
{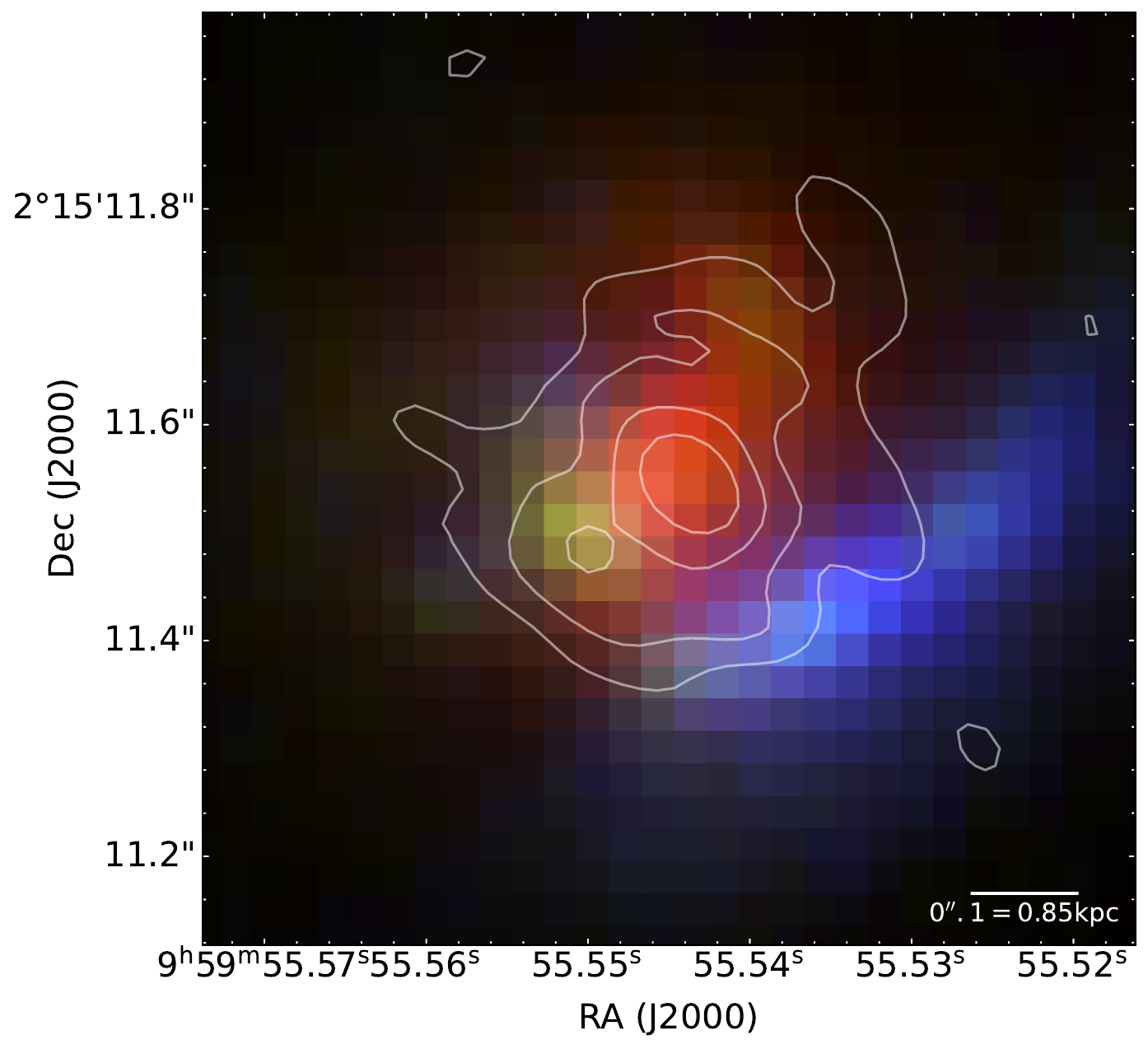}
\includegraphics[width=8cm]{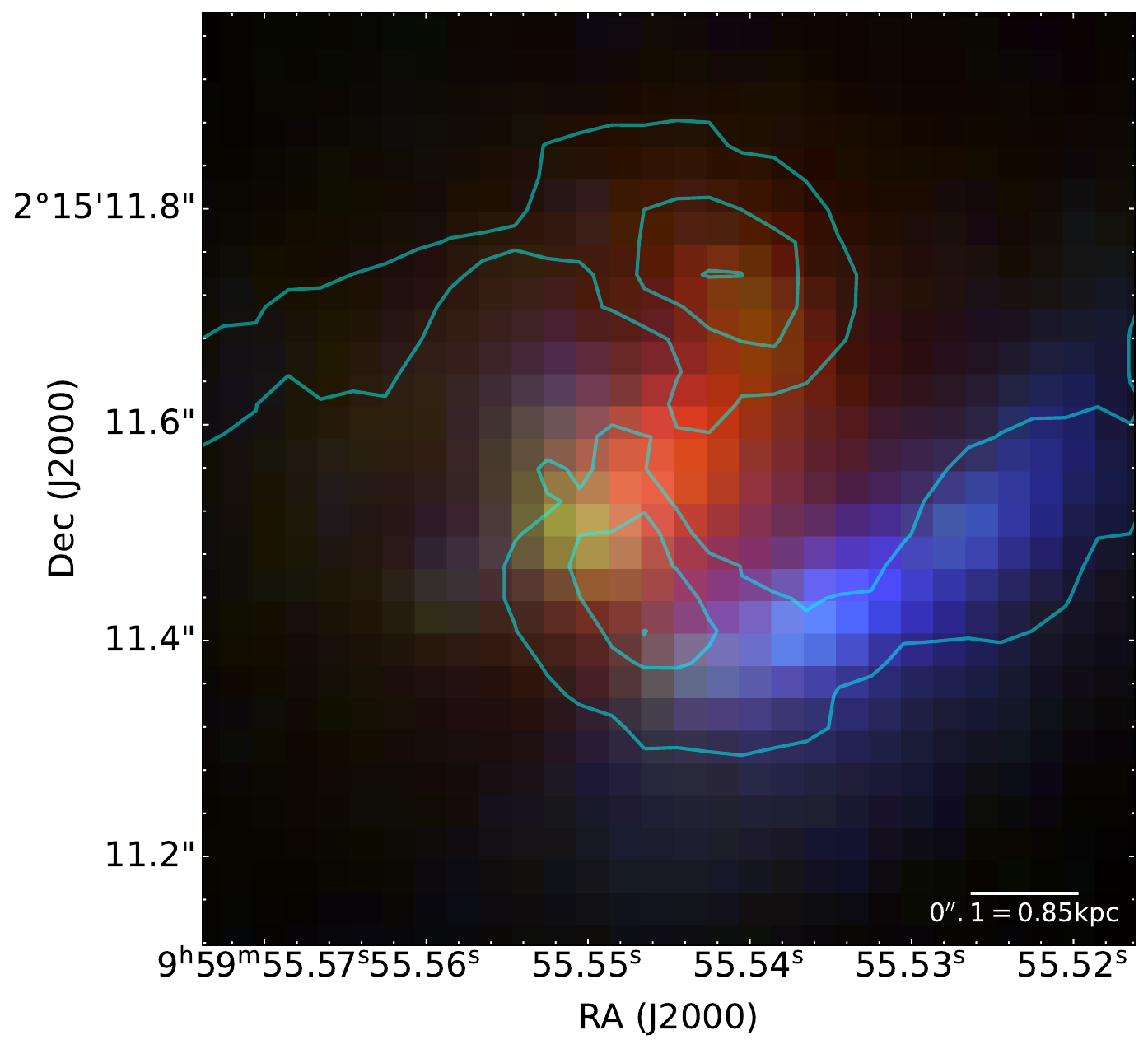}
\caption{RGB image of PACS-819 constructed using the F814W (blue), F150W (green), and  F277W (red) images. $Top$: The ALMA CO (J=5--4) emission is shown by the overlaid contours. Three components are evident: a 'red' dusty central starburst, a dust-free 'blue' emitting component and a fainter source (in 'green') associated with a secondary peak in the CO J=5--4 emission. The contour levels are at 3, 5, 8, 11, 15 $\times \sigma_{\mathrm{rms}}$, where $\sigma_{\mathrm{rms}}=0.037\,\mathrm{Jy}\, \mathrm{beam}^{-1}\,\mathrm{km}\,\mathrm{s}^{-1}$. $Bottom$: The residual structure (Figure~\ref{fig:img_model_fit}) is shown by the cyan contours. The lower spiral arm feature is co-spatial with the 'green' and 'blue' components while a northern structure peaks close to another faint peak of 'green' emission.}
\label{fig:rgb}
\end{figure}

\section{Results: I. The structural nature of PACS-819} \label{sec:results}

The mode (i.e., major merger or secular process) through which starburst activity at high redshift ($z\gtrsim1$) occurs has been previously hampered by the lack of rest-frame optical and near-infrared imaging at sufficient spatial resolution. The smooth nature of the gas and dust in high-redshift galaxies \citep[e.g.,][]{Tadaki2020} imaged by ALMA, has been at odds with rest-frame UV imaging from HST as is the case for PACS-819 (Figure~\ref{fig:obs819}). Here, we present results combining the capabilities of both JWST and ALMA on spatially-resolved scales, required to discern its intrinsic nature.

\begin{figure}
\plotone{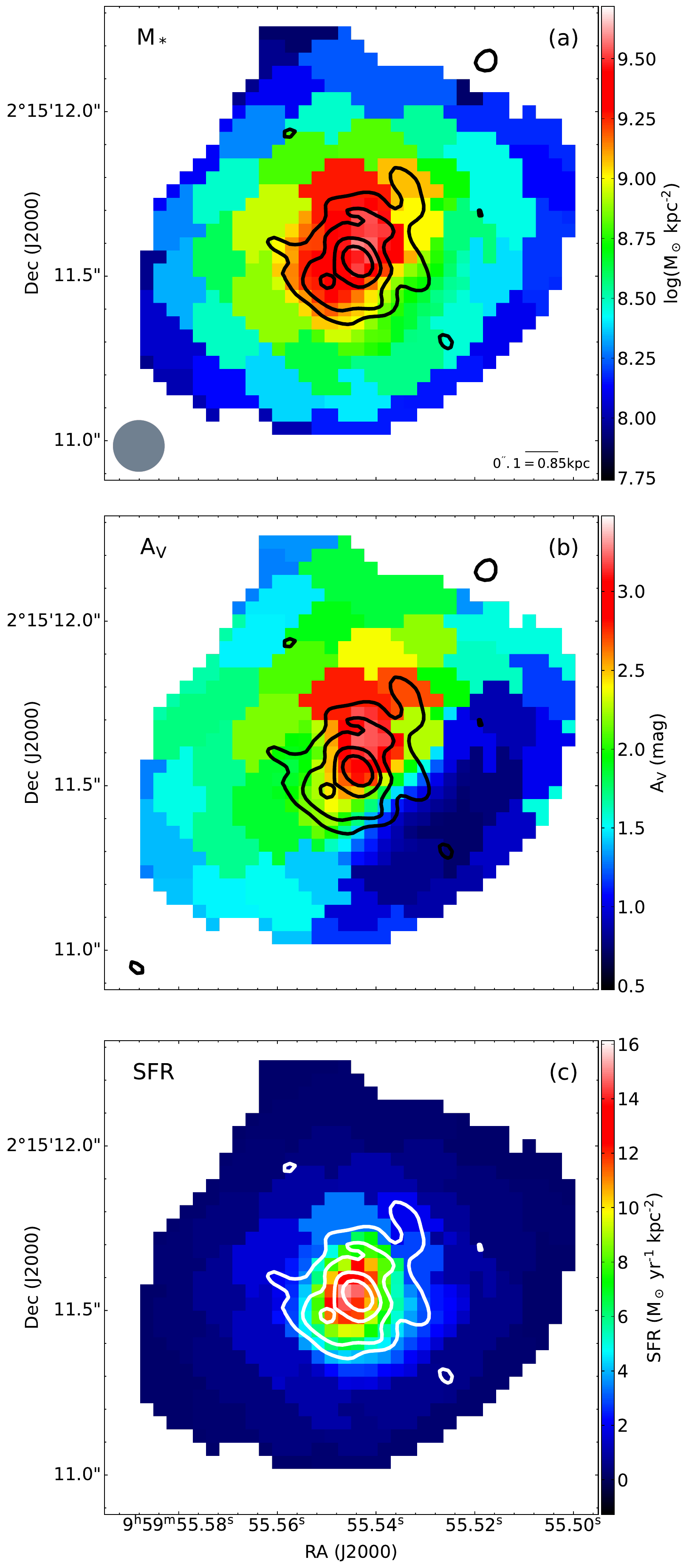}
\caption{Spatially-resolved SED fitting: stellar mass (a), dust extinction (b), and SFR (c). CO J=5--4 contours are overlaid for comparison. 
\label{fig:p2p}}
\end{figure}

\subsection{Unveiling the stellar components with JWST}

In Figure~\ref{fig:obs819}, we present the JWST NIRCam and MIRI images of PACS-819 as provided by the COSMOS-Web survey \citep{casey_cosmos-web_2023}. PACS-819 was first described in \citet{silverman_higher_2015} as a galaxy merger based on multiple rest-frame UV components seen in the HST/ACS F814W image, shown in Figure~\ref{fig:obs819}a. With JWST, the images clearly show the morphology changing significantly with wavelengths from appearing to be a merger at bluer wavelengths (panels b--c, rest-frame optical), to displaying normal galaxy features at redder wavelengths (rest-frame near-IR), particularly the F277W (panel d) and F444W (panel e) filters. The longer wavelength images include a central mass concentration and symmetric spiral-like features. Such features are similar to local Sa galaxies as opposed to disturbed tidal tails as expected in a major merger, especially for a starburst at $11\times$ above the MS as determined from broad-band photometry in COSMOS including $Herschel$.

To further discern the structures seen in the F444W image, we fit the galaxy with a single S\'ersic profile using Galight \citep{2020ApJ...888...37D} to remove the smooth, symmetric, and dominant component to the emission. In Figure~\ref{fig:img_model_fit}, we show the results from this exercise. The main galaxy has a S\'ersic index of 1.5 and a half-light radius of 1.52 kpc. The residual map demonstrates that the remaining structure is dominated by spiral-like features. Therefore, we conclude that the structures seen in the redder JWST images are indicative of a typical star-forming galaxy with no strong signatures of being in an ongoing major merger. We note that the wider field-of-view, compared to Figure~\ref{fig:obs819}, shows nearby neighbors which may have had some interaction with PACS-819 (see Discussion and Appendix).

As expected for star-forming galaxies at $z>1$, the galaxy morphology is wavelength-dependent and highly affected by dust obscuration. To illustrate the obscured nature of PACS-819, we combine the F814W, F150W, and F277W filters into an RGB image to represent the blue (rest-frame UV, 3300\AA, tracing the unobscured stars), green (rest-frame optical, 6100\AA), and red (rest-frame near-infrared, 11000\AA, tracing the obscured stellar population) components. As shown in Figure~\ref{fig:rgb}, we identify three primary components: a central highly obscured galaxy/nucleus in red, a bright blue arc (i.e., a disrupted galaxy or, less likely, a spiral arm or tidal tail based on its stellar mass - see below) to the southwest, and a faint secondary feature to the southeast displayed in green. This structure may be an accreted satellite galaxy or a clump which is present as a consequence of a violent disk instability.

\subsubsection{Spatially-resolved SED fitting and stellar masses of each structure}
\label{text:mass_struct}
The nature of the three distinct components, seen in Figure~\ref{fig:rgb} is revealed through spatially-resolved SED fitting as described in Sec.~\ref{sec:sed}. In Figure~\ref{fig:p2p}, we present the spatially-resolved maps of (a) stellar mass ($M_*$), (b) $A_V$, (c) and SFR. The spatial resolution ($\sim$0.16 arcsec) is comparable to the ALMA data. In contrast to the images at shorter wavelengths (e.g., HST-ACS/F814W, JWST-NIRCam/F115W; rest-frame $\sim$3300--4500\AA) that are clumpy and peak at the southwest arc, the stellar mass distribution is fairly smooth, uniform, and peaks ($\sim 10^{9.5}$ M$_{\odot}$ kpc$^{-2}$) in the central region which is devoid of emission in the HST images. The sum of the stellar mass map is in agreement with that reported previously ($10^{10.7 \pm 0.1}~M_{\odot}$; \citealt{liu_co_2021}). The SFR and extinction ($A_V\sim3$) are both highest at the center of the mass map. As a result, the primary galaxy is a central starburst enshrouded in dust (and gas; see~Sec.~\ref{subsec:sr}) which encompasses the majority of the stellar mass. 

We gauge the importance of the 'blue' and 'green' features by assessing their stellar mass separately. We adopt the method described in \citet{kalita_rest-frame_2023} to determine the mass of the substructures with aperture SED fitting. Segmentation maps are constructed from the F115W image. 
We first convolve the image with a 2D Gaussian kernel using an FWHM of 3 pixels and subtract it from the raw image to highlight the clump features. We then set a 3$\sigma$ detection threshold and perform source deblending with the minimum number of connected pixels set to 5. Using  \texttt{Photutils} \citep{bradley_astropyphotutils_2022}, we subtract the ‘background’ contribution from the 'blue' and 'green' components with consideration of the contribution from the main and central massive galaxy. Our background estimate is based on annular regions with the two clumps being masked.
Since the 'green' component is rather small ($\sim$1kpc) and close to the central main galaxy, we set a 1-pixel wide annulus located on the edge of the clump to estimate a lower-limit to the background contribution thus estimating an upper-limit for the flux of the 'green' component. The error on the flux could be large due to the difficulty in accurately assessing the background. Therefore, we set the error to be 10$\%$ of the flux before background subtraction. For the 'blue' arc, we perform a similar removal of the background and set the flux error to be 10\% of the measured value.

\begin{figure}[h]
\centering
\includegraphics[width=8.2cm]{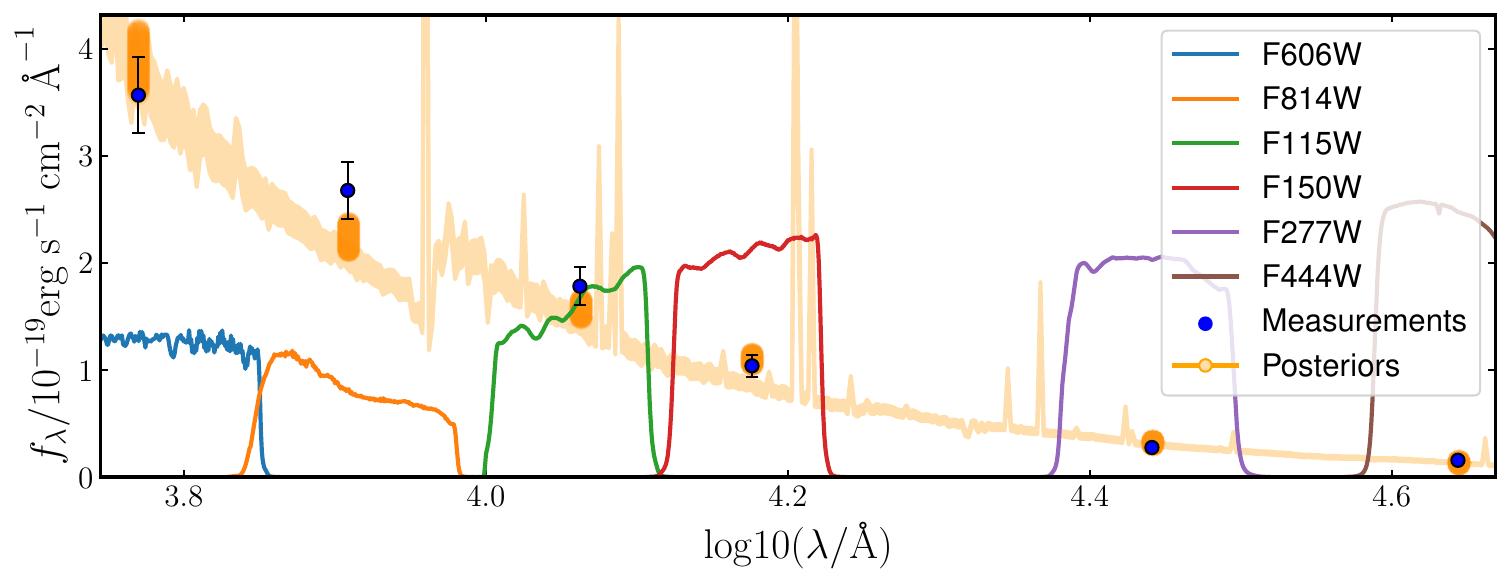}
\includegraphics[width=8.2cm]{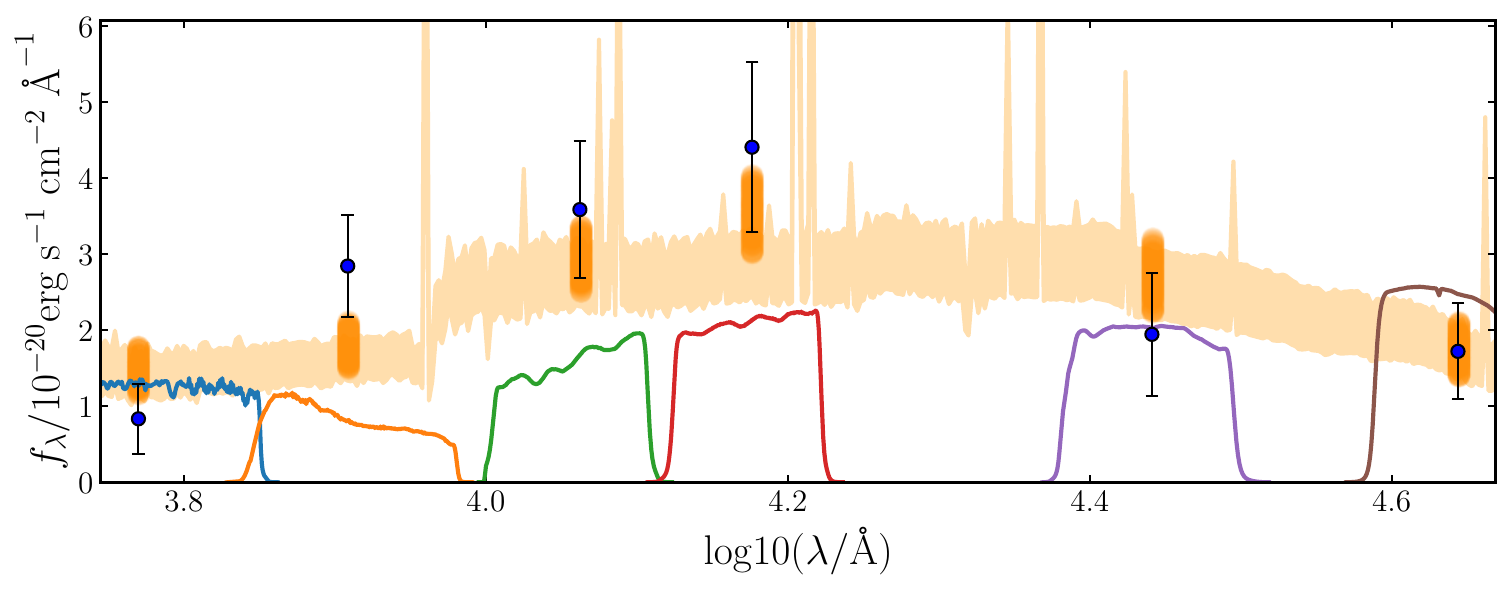}
\caption{SED fitting of the offset HST/JWST structures in PACS-819: $top$: blue arc, $bottom$: 'green' clump feature. Dark points are measurements based on aperture photometry while the orange points and lines are the posterior solution for most likely template. The response profiles of each filter are shown in color with an arbitrary scale.}
\label{fig:sed_examples}
\end{figure}

We repeat the SED fitting (Sec.~\ref{sec:sed}), using the photometry within the segmentation map of each structure and employ a constant SFH model. In Figure~\ref{fig:sed_examples}, we show the SEDs for the 'blue' and 'green' features with best-fit stellar population models. First, the blue arc (to the southwest) is less massive ($log~M_*=8.81_{-0.13}^{+0.12}$, $\sim$1\% of the total mass), hardly noticeable in the mass map (Figure~\ref{fig:p2p}), and has relatively lower obscuration ($A_V\sim 0.2$). Given its mass, this structure may be a tidally-distorted dwarf galaxy, which does not significantly alter the morphology of the central massive galaxy. The stellar mass of this 'green' feature is estimated to be $10^{9.18_{-0.24}^{+0.23}}~M_{\odot}$. However, we conservatively subtracted the background of the 'green' component that may lead to an overestimate of the mass. Therefore, we claim that the 'green' feature constitutes at most $\sim$5\% of the total mass. Each of these three components, spatially offset and differing in color, is fully explored below in the context of the co-spatial gas and dust emission from ALMA.

\begin{figure}[h]
\centering
\includegraphics[width=8cm]{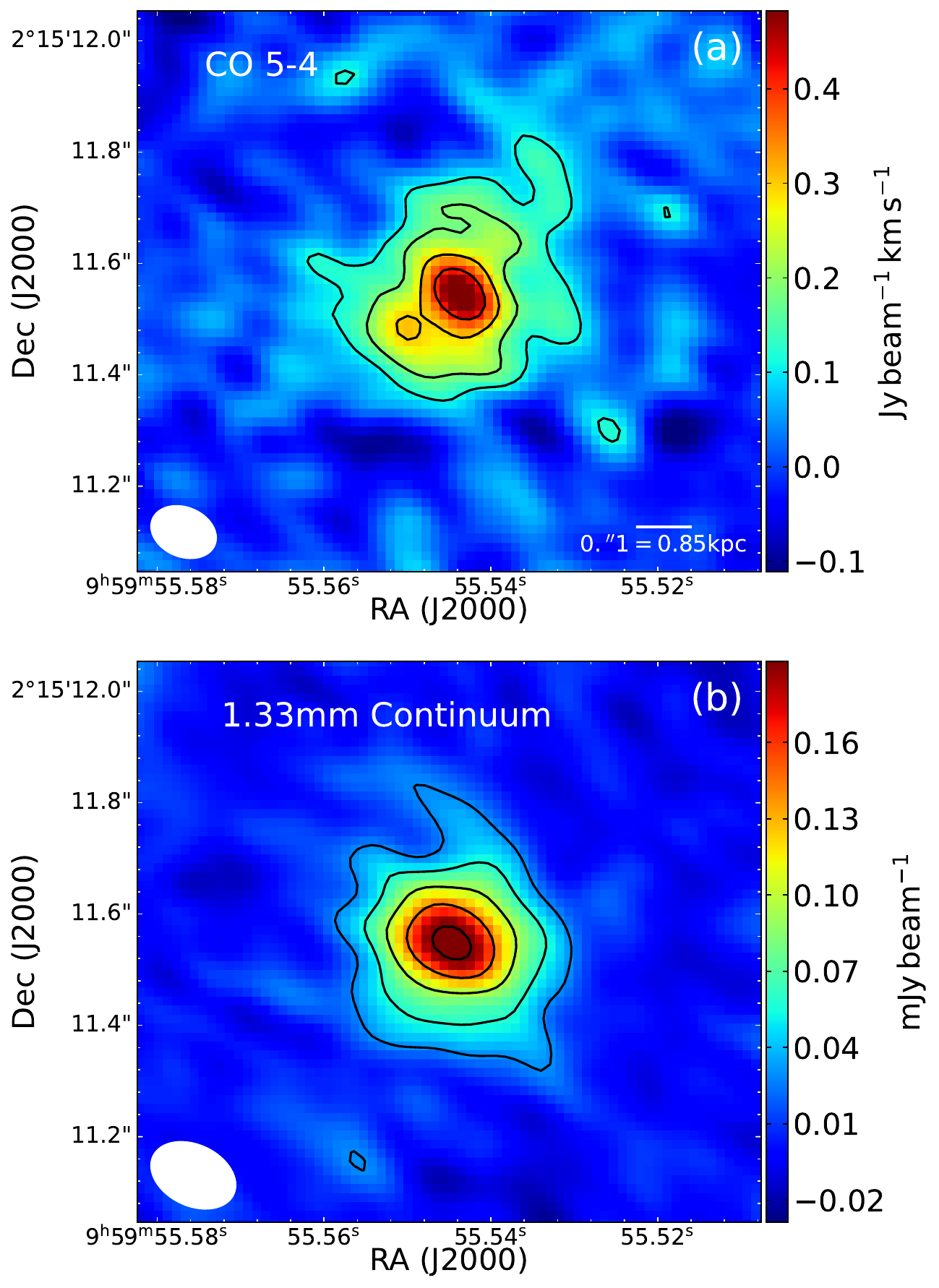}
\caption{ALMA Observations of PACS-819. The panels (a, b) are the CO (J=5--4) and 1.3mm continuum maps at high-resolution. The ALMA beam size is shown in the lower-left corner. The color bars are shown on the right. The black contours in panel (a) are in the same levels as in Figure~\ref{fig:rgb}. Contours in panel (b) are generated from the dust continuum with the levels at 3, 6, 10, 15, 21 $\times \sigma_{\mathrm{rms}}$, where $\sigma_{\mathrm{rms}}\,=\,0.01\,\mathrm{mJy}\, \mathrm{beam}^{-1}$.} For each panel, the north is up and the east is to the left.
\label{fig:obsalma}
\end{figure}

\begin{figure*}
\gridline{\fig{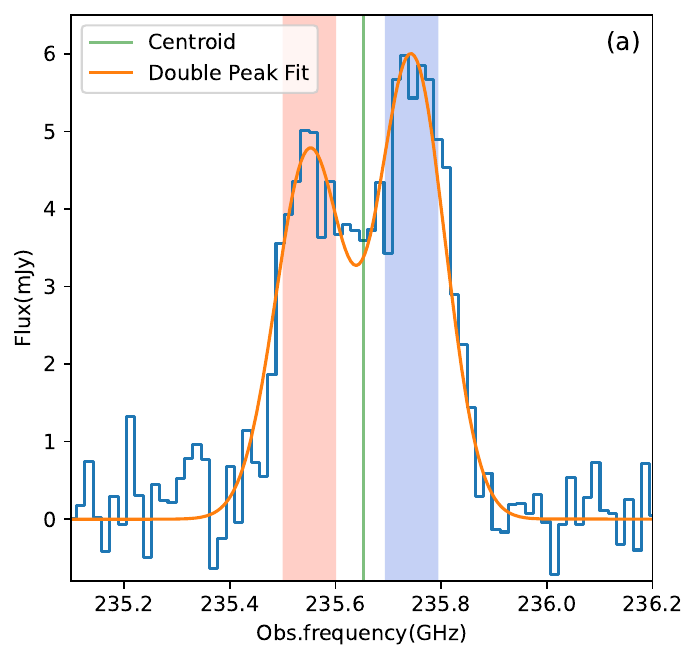}{0.45\textwidth}{}
\fig{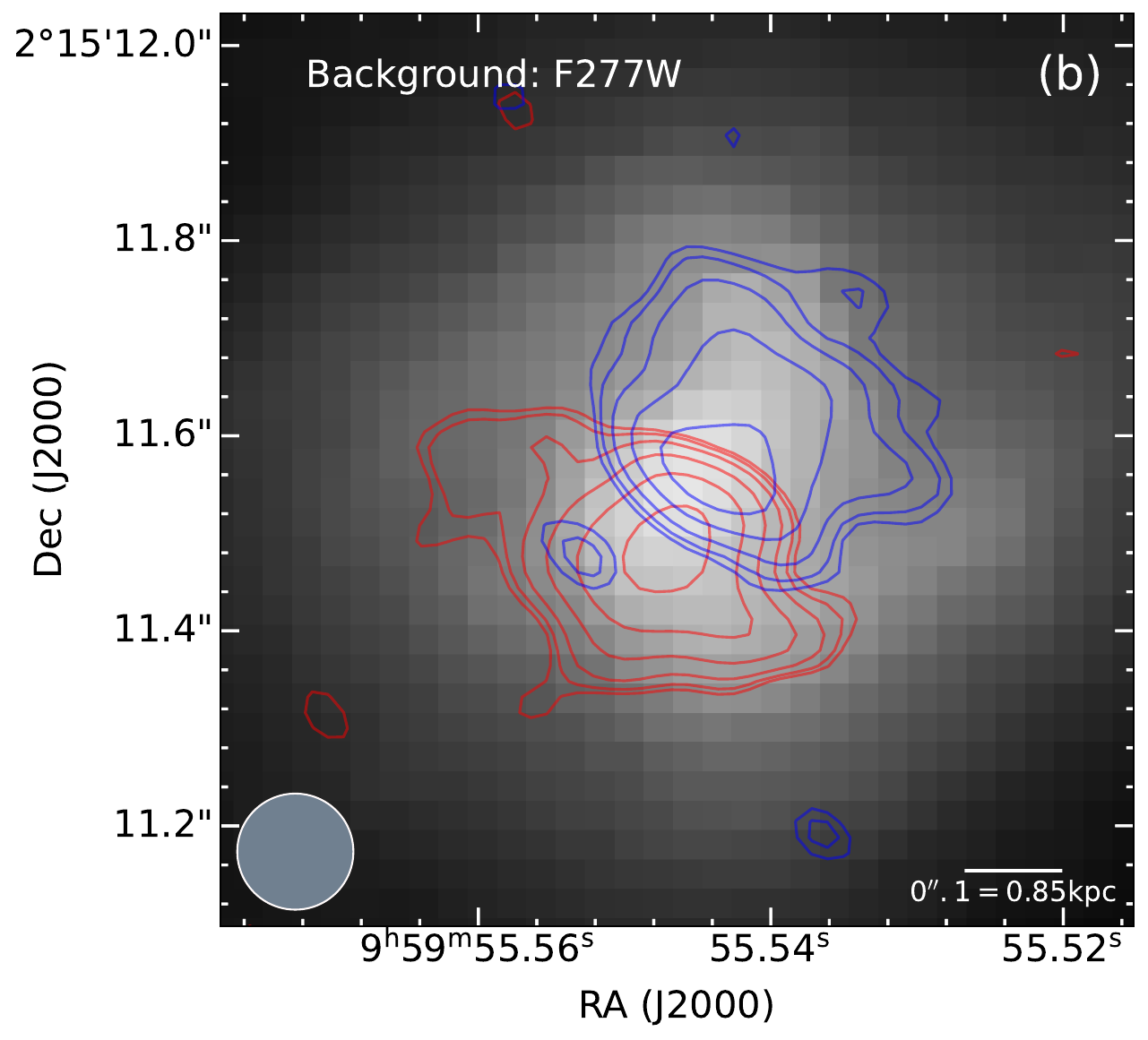}{0.45\textwidth}{}
          }
\caption{(a) CO (J=5--4) spectrum of PACS-819 (blue step function). The centroid is displayed by the green line. The orange line traces our double peak fit result. The red and blue shaded regions represent the intervals centered on the two peaks for kinematics analysis, with widths of $1.6\times \mathrm{std}$, which are 80.6 $\text{km}\,\text{s}^{-1}$ and 80.3 $\text{km}\,\text{s}^{-1}$, respectively. (b) Non-parametric kinematic analysis of PACS-819. The background is the JWST F277W. The red and blue contours are generated from the CO (J=5--4) emission map integrated with redward and blueward intervals shown in panel (a). The contour levels are at 2.5, 3, 4, 6, 9, 13 $\times \sigma_{\mathrm{rms}}$. The $\sigma_{\mathrm{rms}}$ of the red and blue contours are 0.012 and 0.019 Jy/beam $\text{km}\,\text{s}^{-1}$, respectively. A distinct spatial and kinematic structure (blue spot indictated by two low-level contours) is visible above the redshifted part of the disk, co-spatial with the 'green' JWST component seen in the top panel of Figure~\ref{fig:rgb}.
\label{fig:spec}}
\end{figure*}

\subsection{Galaxy kinematics with ALMA using CO J=5--4} \label{subsec:kine}

In the distant Universe, typical star-forming galaxies exhibit disk rotation \citep[see][for a review]{forster_schreiber_star-forming_2020}. However, in the case of extreme starbursts, the kinematic features may differ due to factors such as strong turbulence or merging events that can disrupt ordered disk rotation. To investigate the existence of potential disk features and identify any subcomponents contributing to the enhanced star formation, we began with an analysis of the kinematics by examining the CO (J=5--4) spectrum and producing maps of the emission, both blue- and red-shifted from the line centroid. Following this, we use the 3D-Based Analysis of Rotating Objects via Line Observations ($^{3\mathrm{D}}$BAROLO, \citealt{teodoro_3dbarolo_2015}) package to exploit the high spectral resolution (20 $\text{km}\,\text{s}^{-1}$) ALMA CO (J=5--4) observations at high signal-to-noise.

First, we display in Figure~\ref{fig:obsalma} the ALMA image (panel a) of the full CO (J=5--4) spatial profile, along with the continuum (panel b). The CO emission is detected at high significance and evidently spatially-extended. The peak of the CO emission is centered on the heavily obscured nucleus as shown in the top panel of Figure~\ref{fig:rgb} while a weaker secondary peak is co-spatial with the fainter 'green' component in the RGB image.

We then employ a non-parametric approach to study the kinematics. For spectral analysis, we utilized the \texttt{specflux} task in CASA to extract spectra centered on the CO (J=5--4) emission line, as depicted in Figure~\ref{fig:spec}. The spectrum exhibits a double-peak profile, characteristic of a rotating disk, which is well-fit with a Gaussian model. The FWHMs of the double-peak profiles for PACS-819 are 190.0 $\text{km}\,\text{s}^{-1}$ and 189.2 $\text{km}\,\text{s}^{-1}$ for the low- and high-frequency wings, respectively. The empirical FWHM of the entire line region was found to be 417.7 $\text{km}\,\text{s}^{-1}$. When fitting the profile, we used the mean, amplitude, and standard deviation as initial estimates from CASA viewer, and employed the \texttt{models} function in the \texttt{astropy} package to fit the Gaussian models, allowing all parameters to vary.

We use the \texttt{spectral-cube} python package to produce two moment 0 maps by collapsing data over both redward and blueward narrow intervals, as depicted as the shaded area in Figure~\ref{fig:spec}(a). These intervals were set to be 1.6 $\times$ the standard deviation (80.3 and 80.6 $\text{km}\,\text{s}^{-1}$, respectively) to minimize manual errors. The contours of these collapsed maps, plotted at several $\sigma_{\mathrm{rms}}$ levels, are shown in Figure~\ref{fig:spec}(b) on the F277W image, allowing for a side-by-side comparison of different component profiles.

 The CO kinematics may be indicative of a rotating disk (Figure~\ref{fig:spec}b) as shown by the spatially offset red and blue velocity components. These components are symmetric around the central stellar mass concentration as indicated by the JWST F277W image. There appears to be faint extensions to the CO emission associated with the spiral features; these are seen in the lowest levels of the red contours to the southwest and in the blue contours to the north. Both blue and red contours have considerable extent along the minor axis (northeast to southwest direction), pointing to the possibility of disturbances within the disk.

\begin{figure}
\centering
\includegraphics[width=8cm]{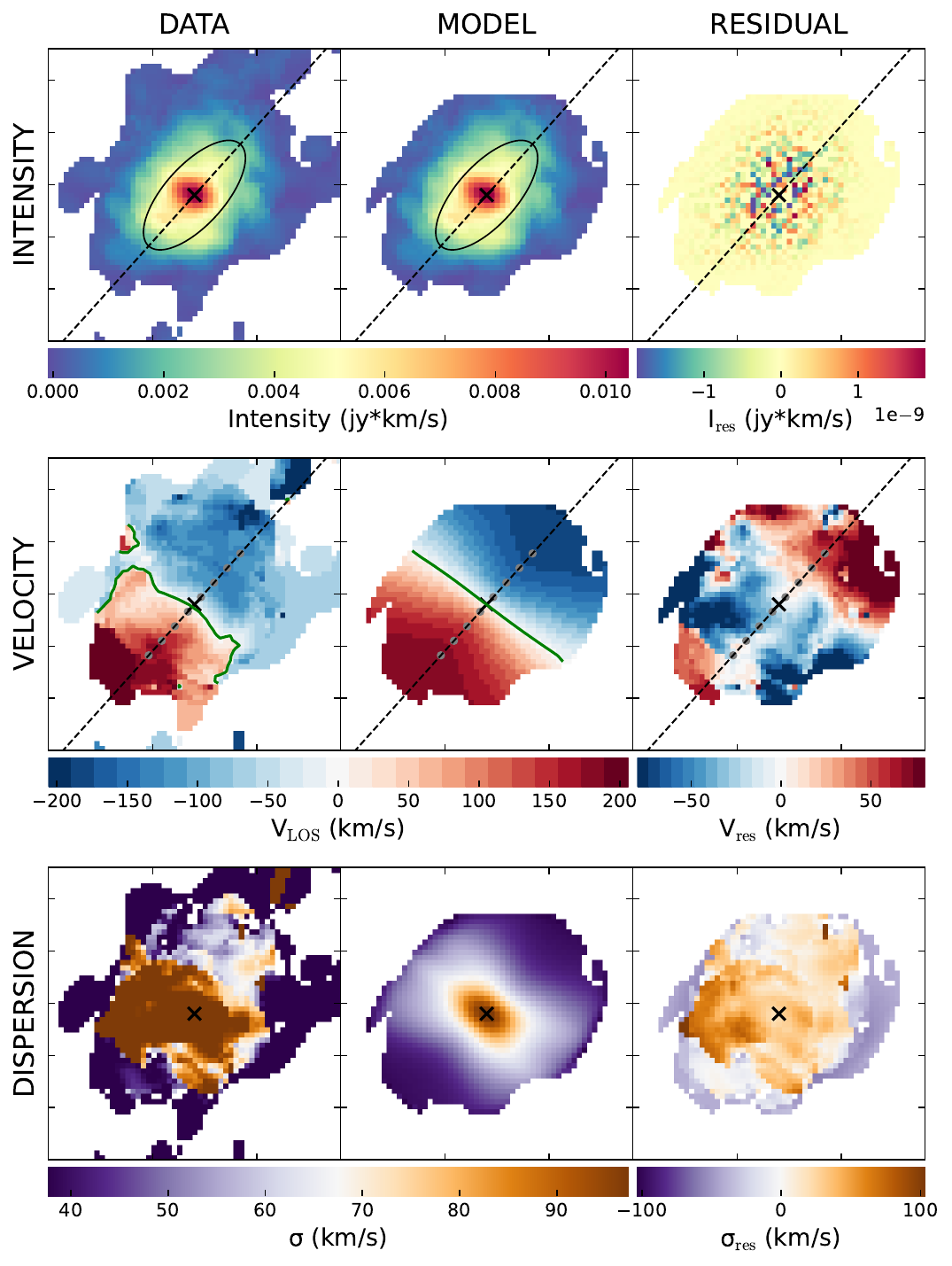}
\caption{$^{3\mathrm{D}}$BAROLO fitting results. From top to bottom are the intensity (moment 0), velocity (moment 1), and dispersion (moment 2) maps. From left to right are the data, fitted model, and residual maps. The black solid circles and crosses mark the half radius and central position of the model while the dashed line marks the position angle of the kinematics major axis. The green line traces the kinematic center where the velocity is zero.
\label{fig:barolo_a}}
\end{figure}

\begin{figure}
\centering
\includegraphics[width=8cm]{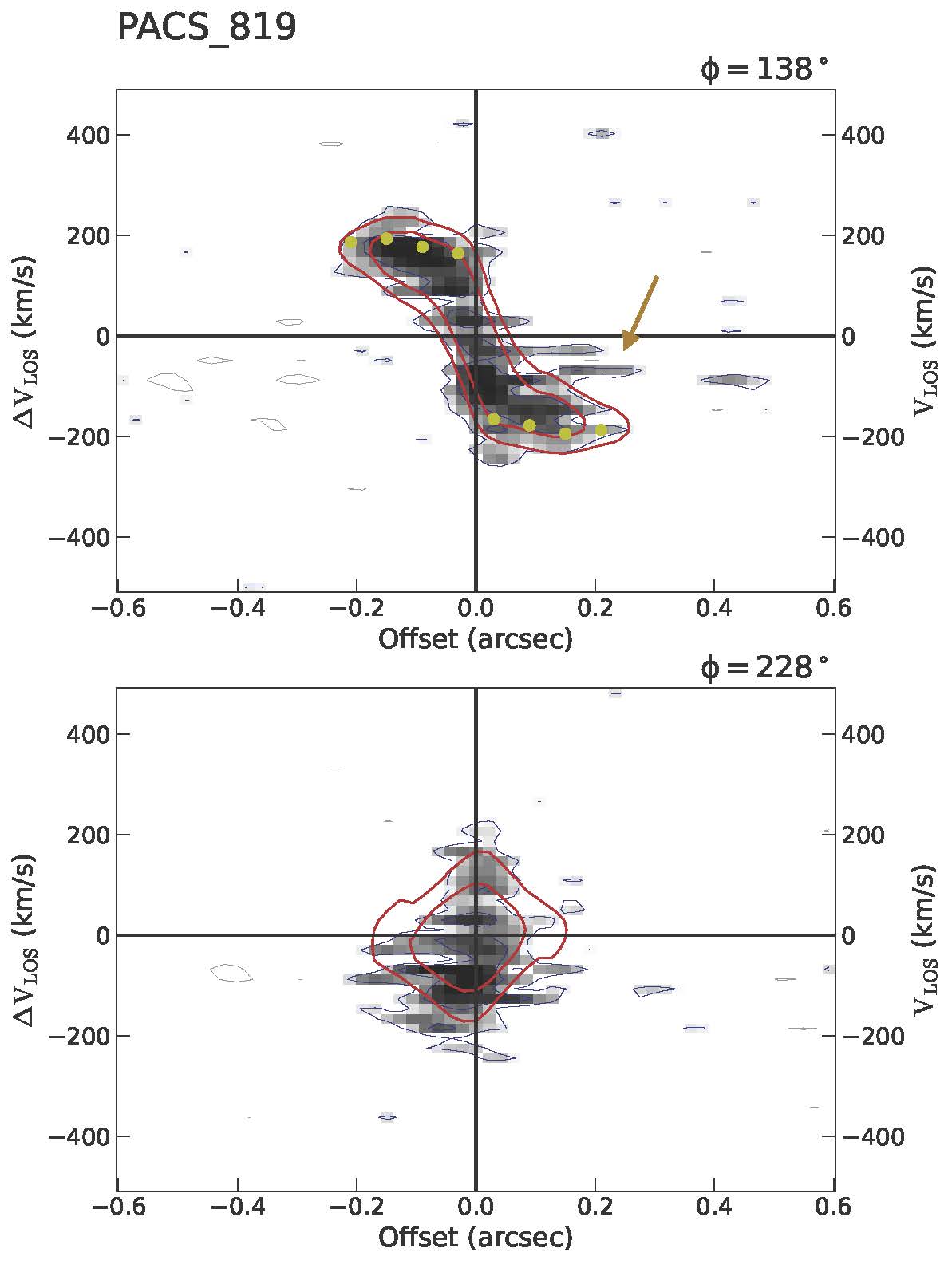}
\caption{Position-velocity map of the CO (J=5--4) emission of PACS-819 along the major and minor axes generated by $^{3\mathrm{D}}$BAROLO. The yellow points on the top panel denote the measurements. The red and blue contours are generated from the model and data cubes, respectively. An arrow highlights the clump feature.
\label{fig:barolo_b}}
\end{figure}

\subsubsection{Model kinematic fitting} 

To determine whether the observed gas kinematics is truly characteristic of a typical rotating disk, we conduct a parametric kinematic analysis of the data cube with the software, $^{3\mathrm{D}}$BAROLO. This software fits the gas with a tilted ring model in 3D, allowing for a better interpretation of the kinematics of the gas, even with just a few resolved components. We successfully build a 3D disk model, as shown in Figure~\ref{fig:barolo_a} with the CO intensity, velocity dispersion given in each column. The rows provide the science data, best-fit model and residual image.

Specifically, we used the SEARCH method to mask the data cube and set the normalization as LOCAL to generate an asymmetry model for our clumpy case. We fixed the system velocity to 0 and input the redshift and rest-frame frequency of our emission line. Due to our S/N limitations, we fixed the RA and Dec of the model and set the initial guess of inclination (INC) and position angle (PA). Then, we employed grid searching in the parameter spaces of INC and PA, with each INC varying by 5 degrees and each PA varying by 15 degrees. We fixed PA as 128 degrees, as it was easier to judge from the velocity field, and searched for the best INC, ultimately fixing it as 45 degrees. We set the rotation velocity $V_{rot}$ as completely free and constrained the dispersion $\sigma_{gas}$ to be higher than 20 $\text{km}\,\text{s}^{-1}$, which is our frequency channel width. $^{3\mathrm{D}}$BAROLO successfully modeled the intensity and velocity field, while dispersion was modeled with less precision due to the challenging nature of modeling it accurately. 

Through our analysis, we find clear evidence for bulk rotation resembling a massive disk as shown in the middle panels of Figure~\ref{fig:barolo_a}. The position-velocity maps (Figure~\ref{fig:barolo_b}) show an 'S'-shape curve characteristic of rotation along the major axis. However, there are kinematic anomalies, particularly on the edge of the residual velocity map at lower S/N (Figure~\ref{fig:barolo_a}), which may be due to the lopsidedness of the disk or the low S/N and fewer velocity channels in these regions. Also, there is a distinct anomaly seen in the velocity and dispersion maps at the position of the blueshifted component (Figure~\ref{fig:spec}b) that is cospatial with the 'green' JWST component in the RGB image (Figure~\ref{fig:rgb}) and likely indicating the existence of a distinct 'clump'. \citet{rizzo_alma-alpaka_2023} analyze the velocity anomalies of PACS-819 in detail and conclude that these anomalies are non-circular features rotating faster or slower than the disk. These features can also be spotted on the maps constructed by our independent study, as shown by the arrow (Figure~\ref{fig:barolo_b}).

\begin{figure}[h]
\centering
\includegraphics[width=7cm]{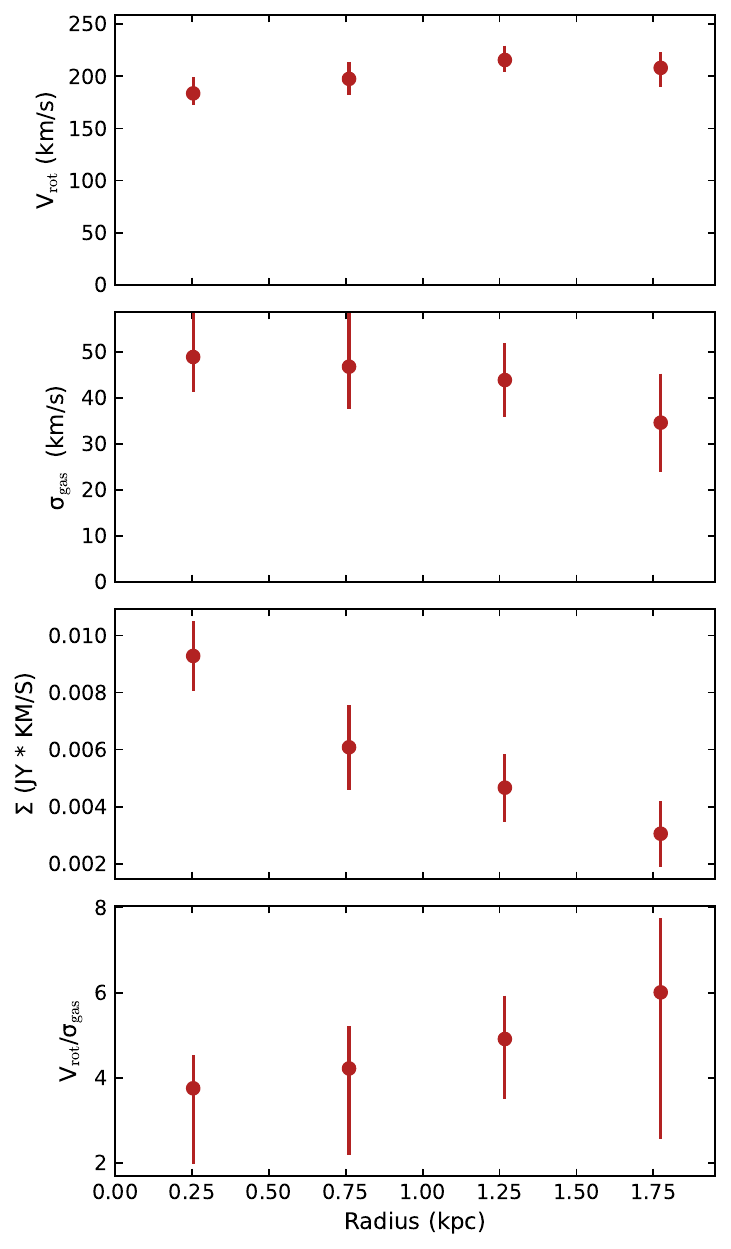}
\caption{Various disk parameters estimated by $^{3\mathrm{D}}$BAROLO, from top to bottom including:
the rotation velocity ($V_{rot}$), dispersion ($\sigma_{gas}$), CO J = 5--4 luminosity surface brightness, and disk stability, as measured by the ratio $V_{rot}/\sigma_{gas}$,
all calculated along the radii.}
\label{fig:parameters}
\end{figure}

We report the measurements of the disk parameters for PACS-819 in Figure~\ref{fig:parameters}. The rotation velocity $V_{rot}$ is about 200 $\text{km}\,\text{s}^{-1}$ and the velocity dispersion $\sigma_{gas}$ is around 40 $\text{km}\,\text{s}^{-1}$, giving a $V_{rot}/\sigma_{gas}$ of around 5, which is rotation-dominated. In agreement with our assessment given above, PACS-819 has kinematic properties broadly consistent with a rotating disk which is typical for high-z star-forming galaxies \citep{schreiber_sinszc-sinf_2018, johnson_kmos_2018, ubler_ionized_2018, wisnioski_kmos3d_2019}. Our conclusions align with the results presented by \citet{rizzo_alma-alpaka_2023}, which characterize this galaxy as rotation-dominated ($V_{rot}/\sigma_{gas}>6-7$) and with notable non-circular features. Such a feature can be spotted in the region pointed out by the brown arrow in Figure~\ref{fig:barolo_b}. The gas there is moving slower than the disk rotation predicted by the model, which is interpreted as inflowing or outflowing extra-planar gas in \citet{rizzo_alma-alpaka_2023}.

\section{Results: II. spatially-resolved properties of the ISM and obscured stellar populations} \label{subsec:sr}

\begin{figure*}
\plotone{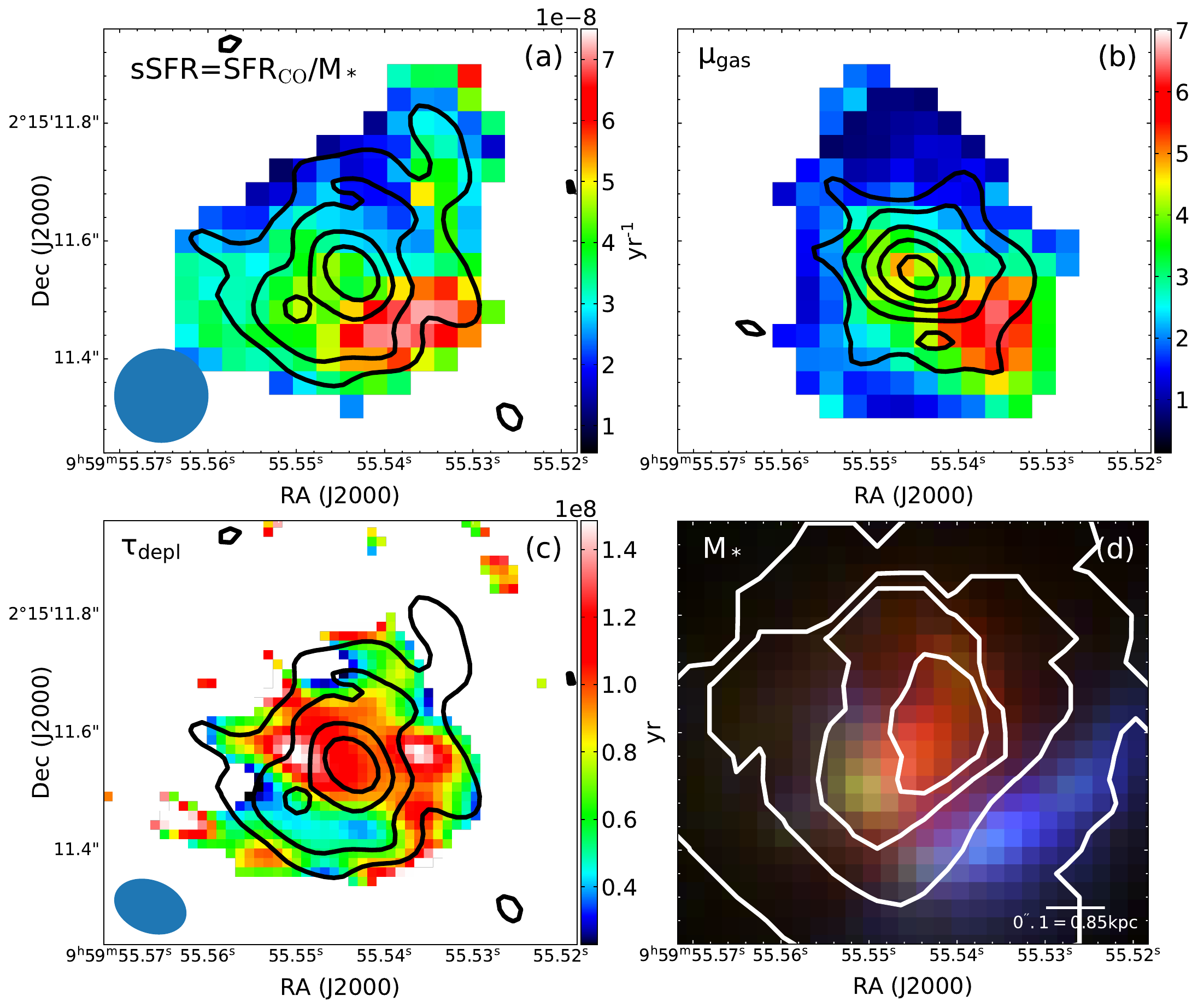}
\caption{Combined ALMA and JWST maps: (a) sSFR (CO/stellar mass; units of yr$^{-1}$), (b) gas fraction ($\mu_{gas}$=$M_{gas}/M_*$), and (c) depletion time ($\tau_{depl}$=M$_{gas}$/SFR; units of yr). The contours in panel b are generated from the dust continuum at the same $\sigma_{\mathrm{rms}}$ levels as CO contours in panels a and c. For comparison, the RGB color map (Figure~\ref{fig:rgb}) with stellar mass contours overlaid is shown in panel d. The levels are manually set at 8.5, 8.9, 9.2, and 9.4 in units of log(M$_\odot$ kpc$^{-2})$.
\label{fig:combined_maps}}
\end{figure*}

Here, we integrate the JWST and ALMA images to generate maps of physical properties as follows: (1) sSFR ($SFR/M_*$) using the CO (J=5--4) emission as a tracer for star formation, (2) gas fraction ($M_{gas}/M_*$) with the dust continuum as a proxy for the gas mass, and (3) the gas depletion time ($t_{depl}=M_{gas}/SFR$). The stellar mass map produced by the JWST and HST images is used for the first two of these. In Figure~\ref{fig:combined_maps}, the maps of these physical properties are shown along with the RGB image and overlaid contours of stellar mass to aid in the visualization. In addition, we examine the location of spatial regions of PACS-819 on the resolved Schmidt-Kennicutt relation (Figure~\ref{fig:ks}).

\subsection{sSFR} \label{subsec:ssfr}

The CO J=5--4 map is used to infer the two-dimensional distribution of SFR based on the relation between CO (J=5--4) luminosity $L'_{CO5-4}$ and the total infrared luminosity $L_{\mathrm{TIR}}$. This is an empirical linear relation that has been well demonstrated \citep{greve_star_2014,daddi_co_2015,liu_high-j_2015,valentino_co_2020}. The CO (J=5--4) luminosity is calculated for each pixel as follows:

\begin{equation}
L_{\mathrm{CO}}^{\prime}=3.25 \times 10^{7} S_{\mathrm{CO}}~\Delta v~\nu_{\mathrm{obs}}^{-2} ~D_{\mathrm{L}}^{2}~(1+z)^{-3} 
\end{equation}

\noindent $S_{\mathrm{CO}}$ and $\Delta v$ are the flux and velocity range of the CO (J=5--4) line, with $I_{\mathrm{CO}5-4}$ based on the collapsed map across the full CO (J=5--4) line. $\nu_{\mathrm{obs}}$ is the redshifted frequency of CO (J=5--4) line. $D_{\mathrm{L}}$ is the luminosity distance. We then convert the CO luminosity to TIR luminosity by the linear relation in \citet{daddi_co_2015}:
\begin{equation}
\log L_{\mathrm{TIR}} / L_{\odot}=\log L_{\mathrm{CO}[5-4]}^{\prime}/(\text{K}~\text{km/s}~\text{pc}^2)+2.52.
\end{equation}

\noindent We note that there is a scatter of 0.2 dex in this relation that we need to take into account here. Using the relation in \citet{kennicutt_star_1998}, we derive a total SFR of 626 $\pm$ 142 $M_{\odot}\text { yr }^{-1}$ with the TIR luminosity as the tracer based on the following calibration by \citet{hao_dust-corrected_2011} and \citet{murphy_calibrating_2011}:
\begin{equation}
\log \mathrm{SFR}\left(M_{\odot} \text { year }^{-1}\right)=\log L_{\mathrm{TIR}}\left(\text{erg/s}\right)-43.41 .
\end{equation}

\noindent Our derived SFR confirms that PACS-819 is a starburst with an SFR elevated by a factor of 11 above the MS \citep{speagle_highly_2014}, given t = 4.32 Gyr (z = 1.45), which agrees with the previous $Herschel$ result as 533$^{+68}_{-60}$ $M_{\odot}\text { yr }^{-1}$. By comparing the total flux with the that ($I_{\mathrm{CO}5-4}$=3.22 $\pm$ 0.18) measured from the low-resolution observation ($\sim0.^{\prime \prime}8$), we confirm there is no significant missing flux (within $1\times\sigma_{rms}$) with a more extended configuration. In subsequent analyses, we mask regions with emission at low S/N ($<3\times \sigma_{rms}$).

Prior to merging the SFR and stellar mass maps, the SFR image is convolved by a Gaussian kernel to match the FWHM based on an empirical PSF model for the JWST/F444W filter using stars within the JWST image. The SFR map is then reprojected to the grid of the stellar mass map. Finally, the resolved sSFR map is simply derived from the division of the SFR map by the stellar mass map.  As evident in Figure~\ref{fig:combined_maps}a, the sSFR map has values around 2--6$\times10^{-8} \,\mathrm{yr}^{-1}$, a timescale typical for starbursts and shorter than MS galaxies ($\sim$Gyr$^{-1}$). These values are a few times higher than the global sSFR ($0.8\times10^{-8} \,\mathrm{yr}^{-1}$). This demonstrates that star formation has a more compact distribution than the stellar mass, indicating that the starburst is happening in the nucleus thus likely growing the bulge. There are variations across the CO-emitting region such as the emission to the southwest which exhibits an enhancement primarily due to the lower stellar mass in that region.

\subsection{Gas fraction} \label{subsec:gasfraction}

To deduce the dust and gas attributes of the starburst, we apply a straightforward blackbody model complemented by conversion factors from existing research. We adopt the black body emission model to infer the dust mass $\mathrm{M_{dust}}$. Utilizing the mass-weighted, average mass-to-light ratio and $\langle U \rangle$=30 estimated by the SED fitting from \citet{liu_co_2021}, we subsequently translate this to a dust temperature $\mathrm{T_{dust}}$ of 33 K, assuming $\langle U \rangle = (\text{T/K} /18.9)^{6.04} $as in \citet{magdis_evolving_2012}. The dust mass is then estimated using the following:

\begin{align}
\label{equation:gas}
M_{\text{dust}} = \frac{S_{\text{obs}} \times D_L^2}{(1 + z) \times B_{\nu_{\text{rest}}}(T) \times k_{\nu_{500\mu \text{m}}}},
\end{align}
\noindent where ${S_{\text{obs}}}$ represents the continuum flux, $B_{\nu_{\text{rest}}}(T)$ is the blackbody function at rest-frame frequency and temperature T, and the empirical dust mass absorption coefficient $k_{\nu_{\text{rest}}}$ at 500 $\mu m$ is 0.051 $\, \text{m}^2/\text{kg}$, as estimated by the Herschel Reference Survey \citep{clark_empirical_2016}.

Given the gas-to-dust ratio of 35 estimated from the study of another starburst of similar redshift (z=1.5) in our survey \citep{silverman_concurrent_2018}, the integrated gas mass is approximated at 4.6 $\pm$ 0.3 $\times 10^{10}M_{\odot}$. This is comparable to the gas mass of $4.6 \times 10^{10}M_{\odot}$ assessed using CO (J=2--1) in \citet{silverman_molecular_2018}, based on a conversion factor $
\alpha_{\mathrm{CO}}=1.3 M_{\odot}/ (\mathrm{K}~\mathrm{km}~\mathrm{s}^{-1}~\mathrm{pc}^2)$.

As with the SFR map, we regrid and smooth the gas map to match the F444W image and then simply divide the gas and stellar maps to produce an image of the gas fraction ($\mu=M_{gas}/M_{*}$). As depicted in Figure~\ref{fig:combined_maps}b, the core region exhibits a high gas fraction peaking at $\sim$3, supportive of a large reservoir of gas supplying the central starburst. The pronounced sSFR, in tandem with the elevated gas fraction, hints at an accelerated stellar mass accumulation in the center. In addition, the peak to the southwest is likely due to the lower stellar mass in that region as also seen in the sSFR map.

\subsection{Resolved KS relation}

With our resolved CO (J=5--4) map serving as the SFR map and the continuum map as a gas mass tracer, we investigate the Schmidt-Kennicutt (KS) law from the inner dense region to the outer regions, generating a resolved KS relation map. The KS law, first introduced by \citet{schmidt_rate_1959} and later refined by \citet{robert_c_kennicutt_global_1998}, describes a power-law relation between the surface densities of SFR and gas mass, providing insights into the efficiency of star formation from gas. In order to reduce the influence of beam size, we use the same weighting factor (0.5) when generating the continuum and CO (J=5--4) maps. Thus, after masking the emission under 2$\times \sigma_{rms}$ and converting the CO (J=5--4) and continuum map into the SFR and gas mass map using the method introduced in Section \ref{subsec:sr}, we derive the resolved map of the gas depletion time (M$_{gas}$/SFR; Figure~\ref{fig:combined_maps}c). We can see various features with slightly differing depletion times. A central elongated region, running almost East to West, of the disk looks uniform at about 120 Myr with two clumps having longer depletion times over 150 Myr. However, we caution that these variations in depletion time are likely to be within the uncertainties and affected by a spatially varying gas-to-dust conversion factor which is beyond the scope of this work.

\begin{figure}[ht]
\centering
\epsscale{1.3}
\plotone{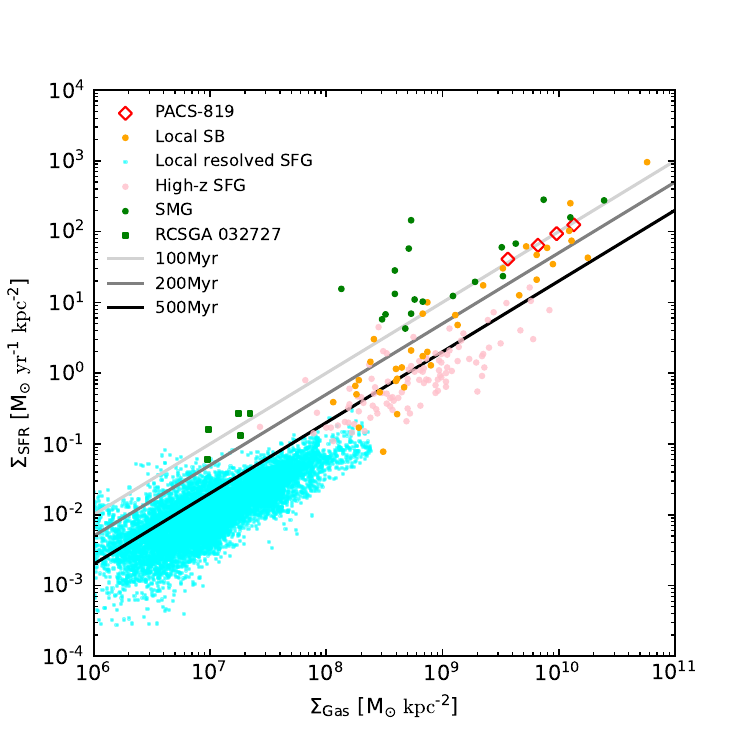}
\caption{SFR density as a function of the molecular gas surface density. The resolved K-S relation measurements from annulus photometry are shown by the red diamonds. For comparison, we display various samples as noted: local starbursts -- yellow dots \citep{robert_c_kennicutt_global_1998}, local resolved star-forming galaxies -- light cyan dots\citep{lin_almaquest_2022}, high-redshift star-forming galaxies -- pink circles \citep{genzel_study_2010,freundlich_towards_2013,tacconi_phibss_2013}, and submillimeter galaxies -- green circles \citep{bothwell_high-resolution_2010,carilli_imaging_2010,genzel_study_2010}. Resolved K-S relation measurements for a young high-z metal-poor starburst \citep{gonzalez-lopez_alma_2017} are shown as small green squares. The 500 Myr gas depletion timescale is delineated by the black line, while the 100 and 200 Myr timescales appear in light gray and gray lines, respectively, aligning closely with the PACS-819 measurements. Uncertainties of the measurements within the map are much smaller than the symbol size.}
\label{fig:ks}
\end{figure}

To display these results on the KS relation (Figure~\ref{fig:ks}), we employ circular annuli with radii initially set to 1-4 times half of the major axis of the synthesized beam's FWHM. In agreement with the above-mentioned depletion times, all our aperture measurements indicate a depletion time around 120 Myr. Therefore, PACS-819 has spatially-resolved depletion times similar to local starbursts and some high-z star-forming galaxies. We observe no significant variation within each annulus, indicating that the disk is forming stars with similar efficiency from the inner to the outer region on average.

\section{Discussion}
We address the different scenarios and processes that can trigger such an extreme starburst event (11$\times$ above the MS) in PACS-819, based on our detailed spatially resolved analysis on the stellar mass, molecular gas and star-forming activity from the multi-wavelength observations with HST, JWST and ALMA.

First, we review the structural nature of PACS-819. The system has three primary components based on their rest-frame colors. A central red component is strongly obscured (i.e., HST-dark) and exhibits intense star-forming activity. This is supported by the $A_V$ map based on spatially-resolved SED fitting (Figure~\ref{fig:p2p}b) and the dust continuum emission seen by ALMA (Figure~\ref{fig:obsalma}b). As demonstrated by the stellar mass distribution (Figure~\ref{fig:p2p}a), this component accounts for majority of the mass and has a rather smooth distribution. The red component exhibits disk-like kinematics, i.e., rotational support with $V_{rot}/\sigma_{gas}\sim4$ based on an analysis using 3D Barolo (Figure~\ref{fig:barolo_b}) with clear kinematic anomalies.

The physical properties of the associated structures are important to discern the nature of the full system. The 'green' component is a less dust-obscured clump evident in both the rest-frame UV/optical and CO (J=5--4) emission with a mass contains up to 5 percent of the entire system. It is kinematically distinct in velocity space and has considerable excess in the dispersion map (Figure~\ref{fig:barolo_a}). The blue component is a prominent extended arc peaking in the rest-frame UV and observed in all HST and JWST NIRCam bands but having no counterpart in the CO (5--4) and dust continuum. This component is almost free of dust and has rather low mass with up to 5 percent of the whole system.
Given the masses of these structures, we estimate that the main obscured 'red' component comprises at least 90\% of the total stellar mass thus ruling out the possibility that these three components are in an ongoing major merger.

\subsection{Minor merger scenario}

By disfavouring a direct major merger event, the next likely process is ongoing minor merging. In support, the rest-frame optical and near-infrared emission is highly asymmetric with multiple peaks even in JWST/F150W (rest-frame $\sim6100$~\AA). The blue arc-like component seems to be a dwarf galaxy tidally-distorted by the massive red component. 
The kinematic anomalies seen in the residual velocity maps may be an indication of torques on the gas due to interactions or past impacts of lower mass galaxies. For example, the 'green' clump may be a recently accreted satellite galaxy which could have contributed to the boost in star formation.

In addition, there is a possibility for the starburst in PACS-819 to be partially induced or assisted by interactions with nearby galaxies. There is spectroscopic evidence for one of the two galaxies to the north to be at a similar redshift to PACS-819 (see the Appendix) thus in close proximity ($\sim$10 kpc away in projection).

\subsection{Secular process}

Alternatively, a more secular process, such as a violent disk instability (VDI; \citealt{dekel_wet_2014,dekel_conditions_2023}) is worth considering given the galaxy morphology, disk rotation, and low mass of the clumps. An in-falling gas stream could introduce a gravitational instability within the disk that can drive gas efficiently to the nuclear region and sustain nuclear starburst activity while keeping the rotating disk configuration intact \citep{dekel_formation_2009,dekel_cold_2009}. Such a scenario may explain the significant increase in the velocity dispersion at the location of the 'green' feature suggestive of turbulence possibly related to gravitational instabilities. The mass of the 'green' clump is similar to those seen in a recent study with JWST \citep{kalita_near-ir_2024}. However, a VDI is not expected to have a velocity offset from the disk itself but rather just show broadening in the dispersion. This lends weight to the likelihood of a past accretion event rather than a VDI.

\section{Conclusions}

In this study, we present JWST and ALMA observations of the distant starburst system PACS-819 (z=1.445), located 11$\times$ above the star-forming MS from $Herschel$ photometry. A unique aspect of our study is the joint use of high-resolution JWST and ALMA observations (approximately 0$^{\prime\prime}$.1 or 0.8 kpc), allowing us to delve into the intricacies of relations between stellar mass, morphology, star formation, dust content, gas properties, and kinematics to address the mechanisms responsible for starburst activity in the gas-rich high-z Universe.

Unexpectedly, we find that PACS-819 has characteristics resembling a more typical star-forming galaxy which is not undergoing a major merger. This is based on the following three observations. 

\begin{itemize}

\item \textbf{Morphology:} JWST NIRCam and MIRI images reveal PACS-819 as having a smooth central mass concentration and spiral-like morphology in the redder bands (e.g., F277W), inconsistent with the emission seen in the bluer bands with JWST and HST.

\item \textbf{Kinematics:} Based on analysis with 3DBarolo, the 3D distribution of the CO J = 5 -- 4 emission exhibits large disk-like rotation with $V_{rot}/\sigma_{gas}\sim5$.

\item \textbf{Disk substructure:} A distinct feature (i.e., clump), associated the disk, is detected by HST, JWST and ALMA. It's moderately-obscured and has a stellar mass of $\lesssim10^{9.2\pm0.2}~M_{\odot}$ (at most $5\%$ of the total; see Sec.~\ref{text:mass_struct}). While there are kinematic anomalies within the disk possibly due to past accretion events or ongoing interactions, this structure does not appear to significantly affect the global kinematics of the rotating disk (see further below). 
\end{itemize}

However, there are spatial and kinematic features that lend favor to the influence of a minor merger event and interactions as listed here.

\begin{itemize}

\item The emission seen in the bluer JWST bands (e.g., F150W; rest-frame $\sim6100$~\AA) is highly asymmetric, has multiple peaks, and includes a prominant blue arc-like feature, as mentioned above, possibly a tidally-distorted dwarf galaxy also undergoing a starburst. 

\item There are kinematic residuals in the CO J=5--4 velocity and dispersion maps which indicate a disturbed disk, possibly induced by neighboring galaxies. The increase in the dispersion near the location of the 'green' structure (i.e., a low-mass satellite galaxy) may be evidence of an impact, particularly given the velocity offset which would not be present if the feature was a VDI.

\end{itemize}

Taken together, the total system is dominated by a nuclear starburst which is building its central bulge through primarily secular processes with assistance from the accretion of a less massive satellite galaxy and interactions with nearby neighbors. This amalgamation of physical effects may illustrate the processes involved in triggering starbursts in the distant gas-rich Universe that lead to further galaxy compaction \citep{tacchella_confinement_2016,kalita_rest-frame_2023}. After the starburst core gradually quenches in an inside-out manner \citep{tacchella_evolution_2016}, PACS-819 may evolve into a clumpy galaxy with a stable disk and substructures, as in UDF2 \citep{rujopakarn_jwst_2023}, and eventually into a massive quiescent galaxy in the nearby Universe.

In addition, we have constructed maps of physical properties (specific star formation rate, gas fraction, depletion time) which further elucidate the nature of the distinct structures present in PACS-819 and show that starburst characteristics (i.e, high gas fraction, short depletion times) persist on smaller physical scales.

\begin{itemize}

\item \textbf{Three major components within the starburst system:} A composite color (RGB) image - encompassing UV, optical, and near-infrared filters, coupled with CO J = 5 -- 4 and FIR observations - uncovers three components: (1) the primary galaxy: massive, dusty, invisible to HST, yet exhibiting strong CO and continuum emission, (2) a secondary clump (as mentioned above): less dusty, of lower mass, aligning with the secondary peak in CO J = 5 -- 4, and dominant in the blue JWST bands, and (3) a UV-bright arc likely representing a tidally-disturbed galaxy.  

\item \textbf{Spatial distribution of the obscured gas, dust and star formation:} The gas and dust distribution are centrally-concentrated mainly within the central kpc ($r_{eff}=1.1\pm0.1$ kpc) based on the FIR continuum image. This is supported by the CO J=5---4 emissions ($r_{eff}=1.2\pm0.1$ kpc), which is known to correlate well with total infrared luminosity and thus ongoing star formation.

\item \textbf{Differential distribution in resolved sSFR and gas fraction maps:} With pixel-to-pixel SED fitting tools and JWST imaging, we construct a resolved stellar mass map to produce that of gas fraction and sSFR. The gas fraction is also centrally concentrated, with a significantly high value $\mu_{\mathrm{gas}}$ (over 3), thus indicating the stellar buildup of the core on rapid timescales (high sSFR), while the integrated gas fraction is $\sim1$, compatible to normal star-forming galaxies at similar redshifts. There is a further enhancement in the gas fraction and sSFR associated with the blue arc (i.e., a tidally-disturbed dwarf galaxy).

\item \textbf{High and uneven distributed SFE:} Based on our resolved analysis, the whole system has a short depletion time of $\sim120$ Myr, aligning with local starbursts and high-z star-forming galaxies. Within the resolved SFE map, there are variations to be further explored.

\end{itemize}

To conclude, PACS-819 is clearly undergoing a nuclear starburst that is likely to be rapidly growing its bulge. Whether such a compact and highly starforming nucleus can occur without a major merger is still an open question. If realized, secular processes would alter our view of the physical drivers of starburst phenomenon in the gas-rich Universe at $z>1$. Most likely, the recent accretion of a low-mass satellite and interactions with neighboring galaxies may be responsible for destabilizing a disk and driving gas to the center thus powering a starburst. In any case, our study underscores the importance of high-resolution multi-wavelength data to further understand the nature and evolution of distant star-forming galaxies.

\begin{deluxetable*}{cccccccc}
\tablenum{1}
\tablecaption{Summary of ALMA observations\label{tab:observations}}
\tablewidth{0pt}
\tablehead{
\colhead{Configuration} & \colhead{Obs. Date} & \colhead{Band} & \colhead{Central Frequency} & \colhead{Band Width} & \colhead{On-source Time} &\colhead{Beam Size\tablenotemark{b}} &\colhead{$\sigma_{\mathrm{rms}}$\tablenotemark{b,c}}\\
 & \colhead{(UT)} & & \colhead{(GHz)} &  \colhead{(GHz)} & \colhead{(minutes)} & \colhead{($^{\prime \prime}$)} & \colhead{(mJy beam$^{-1}$)}
}
\startdata
Compact	&	2016 Dec 16 \& 	& 6	&217.99, 220.99	& 2.000, 2.000& 11.1 + 11.1& 0.83 $\times$ 0.69 &0.02 \\
	&	Apr 13	& 6 	& 233.60, \textbf{235.60}\tablenotemark{a}	&2.000, 1.875 &  & 0.84 $\times$ 0.70 & 0.54\\
Extended	&	2017 Aug 24 \& 	& 6	&217.99, 220.99	& 2.000, 2.000& 35.4 + 16.9 + 35.4& 0.13 $\times$ 0.09 &0.01 \\
	&	Sep 02	& 6 	& 233.60, \textbf{235.60}\tablenotemark{a}	&2.000, 1.875 &  & 0.12 $\times$ 0.09 & 0.41\\
\enddata
\tablenotetext{a}{The central frequency in bold indicates the spectral window where the CO line falls.}
\tablenotetext{b}{The first and second rows are read from the reconstructed continuum and cube with Briggs 0.5 weighting, respectively.}
\tablenotetext{c}{The rms noise of the cube is calculated from the channel where the line is strongest and the widths of the channel are 35 and 20 $\text{km}\,\text{s}^{-1}$.}
\end{deluxetable*}

\begin{deluxetable*}{cccc}
\tablenum{2}
\tablecaption{Measurements and derived properties\label{tab:properties2}}
\tablewidth{0pt}
\tablehead{
\colhead{Quantity} & \colhead{PACS-819} & \colhead{Units} &
\colhead{Notes}
}
\decimalcolnumbers
\startdata
$S_{1.3\text{mm}}$ & 0.639 $\pm$ 0.048& $\mathrm{mJy}$ & total flux of the underlying continuum\\
$r_{\text{cir,CO}}$ & 1.21 $\pm$ 0.11 & kpc & Effective radius of CO (5-4); circularized \\
$r_{\text{cir,con}}$ & 1.08 $\pm$ 0.11 & kpc & Effective radius of continuum; circularized \\
$i_{\text{CO}}$ & $45 \pm 10$& degrees & Disk inclination (CO)\\
$i_{\text{con}}$ & $30 \pm 20$ & degrees & Disk inclination (Continuum)\\
$I_{\mathrm{CO}5-4}$ &2.92 $\pm$ 0.31& $\mathrm{Jy}\,\mathrm{km}\,\mathrm{s}^{-1}$& CO (5-4) intensity\\
$\Delta v$&920.0 &$\mathrm{km}\,\mathrm{s}^{-1}$ &Velocity range over which $I_{\mathrm{CO}5-4}$ is measured\\
$v_{\mathrm{FWHM}}$& 417.7  & $\mathrm{km}\,\mathrm{s}^{-1}$&{CO velocity FWHM}\\
\hline
$M_{\text{ISM}}$ & $4.6\pm0.3\times10^{10}$& $M_{\odot}$ & From 1.3mm continuum\\
$\mu_{\text{gas}}$ & 1.0 $\pm$ 0.1  &  &\\
$L'_{\mathrm{CO}(5-4)}$ & $1.3\pm0.1\times10^{10}$ &$\mathrm{K}\, \mathrm{km}\, \mathrm{s}{ }^{-1} \,\mathrm{pc}^{2}$& CO (5-4) luminosity\\
SFR &626 $\pm$ 142 &$M_{\odot}\, \mathrm{yr}^{-1}$ & From CO (5-4)\\
$\tau_{\mathrm{depl}}$&73 $\pm$ 10 &Myr &\\
\enddata
\end{deluxetable*}

\appendix

\section{Local environment of PACS-819}

In the proximity of PACS-819, there are two galaxies within a 2$^{\prime\prime}$ radius. According to the COSMOS2020 catalog \citep{weaver_cosmos2020_2022}, they are likely foreground galaxies with photometric redshifts as $0.396^{+0.008}_{-0.005}$ and $1.292^{+0.017}_{-0.016}$ estimated with \texttt{EAZY} \citep{brammer_eazy_2008} and $0.405^{+0.010}_{-0.010}$ and $1.222^{+0.051}_{-0.073}$ with \texttt{LePhare} \citep{arnouts_measuring_2002,ilbert_accurate_2006}, based on the comprehensive photometric data set available in COSMOS. In fact, the former galaxy is reported to have a spectroscopic redshift of 1.445 based on a slit spectrum acquired with Keck/DEIMOS \citep{masters_complete_2019} and labeled as a serendipitous detection. After visual inspection, there is no other galaxy detected in CO or the continuum in the ALMA field of view within the observing channels. The non-detection agrees with the properties given in the COSMOS2020 catalog which shows these galaxies might be quiescent ($M_*\sim10^{10}~M_{\odot}$, SFR $<10 M_{\odot}\, \mathrm{yr}^{-1}$).

\begin{figure*}
\plotone{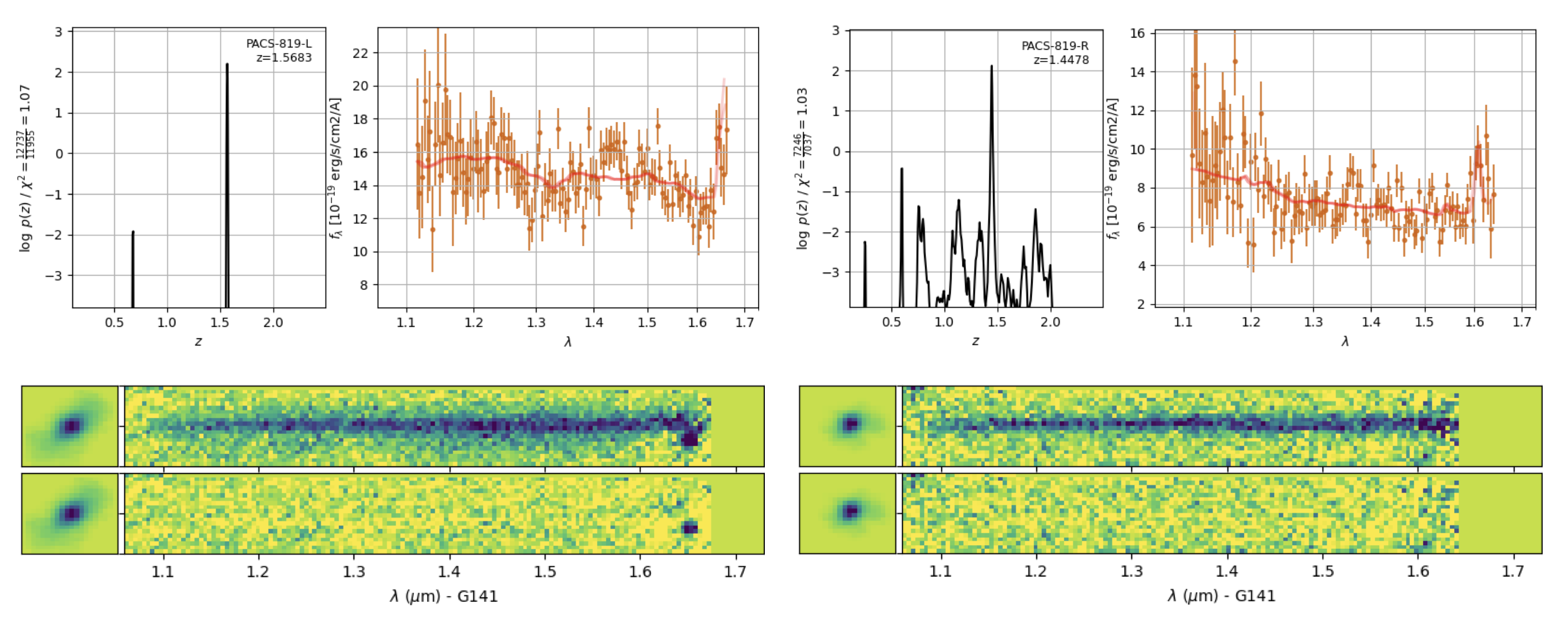}
\epsscale{0.35}
\plotone{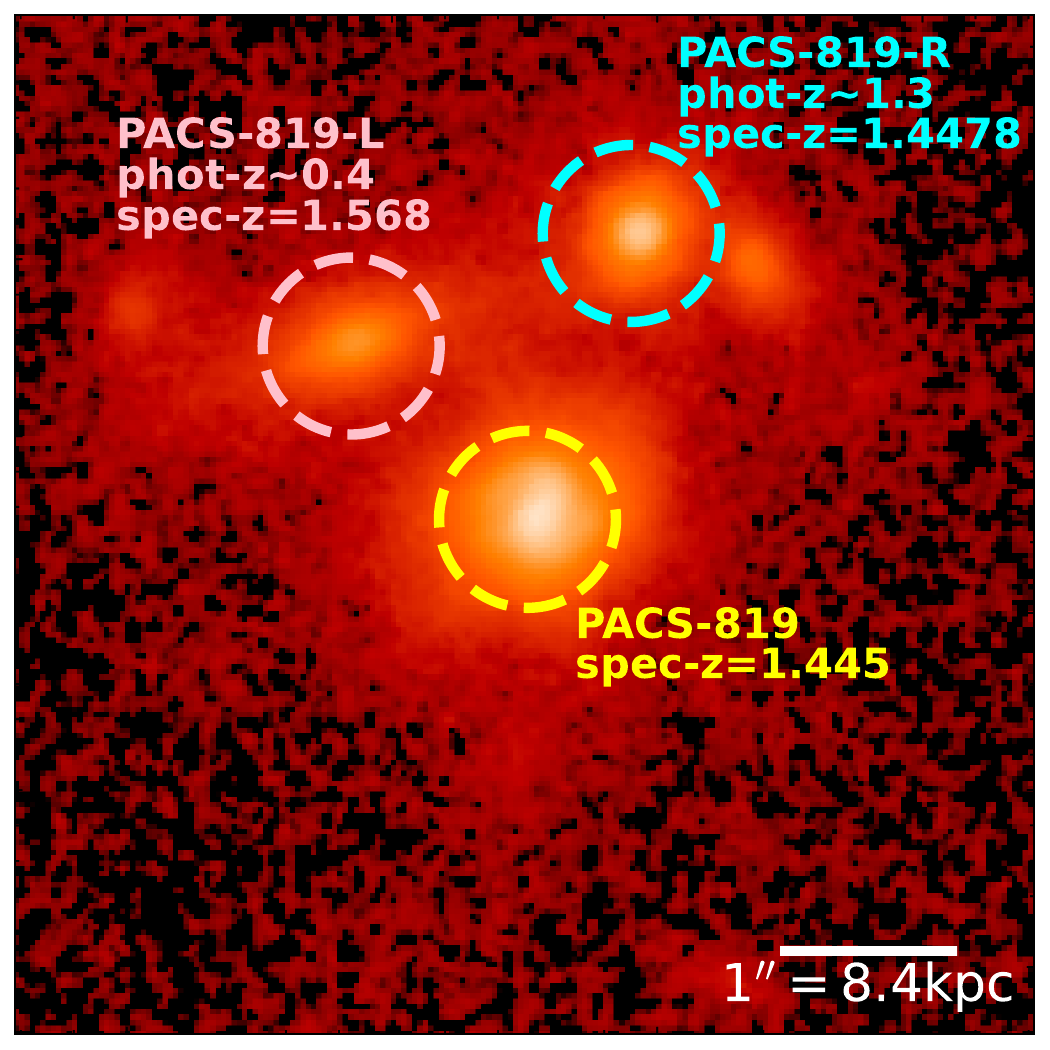}
\caption{HST G141 GRISM spectra and fitting results of the two neighboring galaxies to PACS-819. Top: Stacked spectra (right: orange dots and errors), the probability density function (PDF) of the redshift solution in the left panel, and the best-fit model of the 1D spectra (red line). PACS-819-L refers to the galaxy to the northeast of PACS-819 with a serendipitous detection by Keck. PACS-819-R refers to the galaxy to the northwest. In both spectra, only one significant line can be identified, which is impacted by contamination due to strong H$\alpha$ emission from PACS-819, primarily for PACS-819-L. Middle: 2D spectra of two galaxies. The top row shows the raw data while a continuum component modeled by polynomial is subtracted in the bottom row. The F160W cutouts are plotted on the left of each spectrum. Bottom: JWST F444W image with the galaxies labeled.
\label{fig:grism}}
\end{figure*}

Fortunately, the area around PACS-819 has been observed with HST G141 GRISM (PI: Lemaux; \#16684). We follow the \texttt{Grizli} pipeline \citep{2023zndo...7712834B} to acquire and process the data. The reduced 2D spectra and best-fit models are presented in Figure~\ref{fig:grism}. Unfortunately, we encountered some issues in determining their spectroscopic redshifts when fitting the extracted spectra with \texttt{EAZY}. First, the galaxies are very close (within two arcsecs), so the nearby galaxies are strongly contaminated by the bright H$\alpha$ line emitted by the extreme starburst PACS-819. Second, the H$\alpha$ feature is at the edge of the field of view thus reducing the data quality and making it hard to discriminate any intrinsic spectral features and contamination. For PACS-819-L with a serendipitous detection, the derived spectroscopic redshift ($z=1.568$) is different from PACS-819 ($z=1.445$) even though there may be strong contamination. This disagrees with the spec-z reported for the serendipitous Keck/DEIMOS catalog. In any case, there is no evidence that this galaxy is at an equivalent redshift to PACS-819 to claim that there is an ongoing interation. The fitting result for the other galaxy (PACS-819-R) shows that it has the same spec-z ($z_{spec}=1.4478$) as PACS-819. The $L_{H_{\alpha}}$ from the fit is $9.8\pm1.8\times10^{41}~\mathrm{erg/s/cm^2}$, about two magnitudes lower than PACS-819, indicating this pair might be a dry merger. However, the lineup direction across it and PACS-819 is accidentally perpendicular to the direction of dispersion. Therefore, we can only state that the brighter nearby neighbor is likely to be at the same redshift as PACS-819 thus potentially undergoing an interaction with all members of a galaxy group.

\clearpage

\begin{acknowledgments}

We thank the anonymous referee for their constructive comments. Z.L. sincerely thanks J.S., E.D., and A.R. for their guidance and assistance. Z.L. would extend his sincere gratitude to Frederic Bournaud, Connor Bottrell, Andy Bunker, Y. Sophia Dai, Jing Wang, and Mingyang Zhuang for the scientific discussions and Shenli Tang, Qinyue Fei, Yi Xu, and Fengwu Sun for technical discussions. Z.L. is grateful to Yali Shao and Lihwai Lin for sharing the data points used in Figure~\ref{fig:ks}.

Kavli IPMU was established by World Premier International Research Center Initiative (WPI), MEXT, Japan. Z.L. is supported by the Global Science Graduate Course (GSGC) program of the University of Tokyo. J.S. is supported by JSPS KAKENHI (JP22H01262) and the World Premier International Research Center Initiative (WPI), MEXT, Japan. This work was supported by JSPS Core-to-Core Program (grant number: JPJSCCA20210003). G.E.M acknowledges the Villum Fonden research grant 13160 “Gas to stars, stars to dust: tracing star formation across cosmic time,” grant 37440, “The Hidden Cosmos,” and the Cosmic Dawn Center of Excellence funded by the Danish National Research Foundation under the grant No. 140. AP is supported by an Anniversary Fellowship at University of Southampton. SG acknowledges financial support from the Villum Young Investigator grant 37440 and 13160 and the Cosmic Dawn Center (DAWN), funded by the Danish National Research Foundation (DNRF) under grant No. 140. 

Some of the data products presented herein were retrieved from the Dawn JWST Archive (DJA). DJA is an initiative of the Cosmic Dawn Center, which is funded by the Danish National Research Foundation under grant No. 140.
Some of the data presented in this article were obtained from the Mikulski Archive for Space Telescopes (MAST) at the Space Telescope Science Institute. The specific observations analyzed can be accessed via \dataset[10.17909/9m5v-wk24]{https://doi.org/10.17909/9m5v-wk24}. This paper makes use of the following ALMA data: ADS/JAO.ALMA\#2016.1.01426.S. ALMA is a partnership of ESO (representing its member states), NSF (USA) and NINS (Japan), together with NRC (Canada), MOST and ASIAA (Taiwan), and KASI (Republic of Korea), in cooperation with the Republic of Chile. The Joint ALMA Observatory is operated by ESO, AUI/NRAO and NAOJ.
\end{acknowledgments}

\facilities{HST(STIS), JWST(STIS), ALMA(NRAO, ESO, NAOJ, ASIAA and Chile)}

\software{ $^{3\mathrm{D}}$BAROLO, astropy, bagpipes, CASA, EAZY, Grizli, photutils, pixedfit, spectral-cube, uncertainties.}

\bibliography{ref.bib}{}

\begin{thebibliography}{}
\expandafter\ifx\csname natexlab\endcsname\relax\def\natexlab#1{#1}\fi
\providecommand{\url}[1]{\href{#1}{#1}}
\providecommand{\dodoi}[1]{doi:~\href{http://doi.org/#1}{\nolinkurl{#1}}}
\providecommand{\doeprint}[1]{\href{http://ascl.net/#1}{\nolinkurl{http://ascl.net/#1}}}
\providecommand{\doarXiv}[1]{\href{https://arxiv.org/abs/#1}{\nolinkurl{https://arxiv.org/abs/#1}}}

\bibitem[{{Abdurro'uf} \& Akiyama(2017)}]{abdurrouf_understanding_2017}
{Abdurro'uf}, \& Akiyama, M. 2017, Monthly Notices of the Royal Astronomical Society, 469, 2806, \dodoi{10.1093/mnras/stx936}

\bibitem[{{Abdurro’uf} {et~al.}(2021){Abdurro’uf}, Lin, Wu, \& Akiyama}]{abdurrouf_introducing_2021}
{Abdurro’uf}, Lin, Y.-T., Wu, P.-F., \& Akiyama, M. 2021, ApJS, 254, 15, \dodoi{10.3847/1538-4365/abebe2}

\bibitem[{Arnouts {et~al.}(2002)Arnouts, Moscardini, Vanzella, Colombi, Cristiani, Fontana, Giallongo, Matarrese, \& Saracco}]{arnouts_measuring_2002}
Arnouts, S., Moscardini, L., Vanzella, E., {et~al.} 2002, Monthly Notices of the Royal Astronomical Society, 329, 355, \dodoi{10.1046/j.1365-8711.2002.04988.x}

\bibitem[{Barnes \& Hernquist(1996)}]{barnes_transformations_1996}
Barnes, J.~E., \& Hernquist, L. 1996, ApJ, 471, 115, \dodoi{10.1086/177957}

\bibitem[{{Bertin}(2011)}]{2011ASPC..442..435B}
{Bertin}, E. 2011, in Astronomical Society of the Pacific Conference Series, Vol. 442, Astronomical Data Analysis Software and Systems XX, ed. I.~N. {Evans}, A.~{Accomazzi}, D.~J. {Mink}, \& A.~H. {Rots}, 435

\bibitem[{Bothwell {et~al.}(2010)Bothwell, Chapman, Tacconi, Smail, Ivison, Casey, Bertoldi, Beswick, Biggs, Blain, Cox, Genzel, Greve, Kennicutt, Muxlow, Neri, \& Omont}]{bothwell_high-resolution_2010}
Bothwell, M.~S., Chapman, S.~C., Tacconi, L., {et~al.} 2010, Monthly Notices of the Royal Astronomical Society, 405, 219, \dodoi{10.1111/j.1365-2966.2010.16480.x}

\bibitem[{Bradley {et~al.}(2022)Bradley, Sipőcz, Robitaille, Tollerud, Vinícius, Deil, Barbary, Wilson, Busko, Donath, Günther, Cara, Lim, Meßlinger, Conseil, Bostroem, Droettboom, Bray, Andersen~Bratholm, Barentsen, Craig, Rathi, Pascual, Perren, Georgiev, De~Val-Borro, Kerzendorf, Bach, Quint, \& Souchereau}]{bradley_astropyphotutils_2022}
Bradley, L., Sipőcz, B., Robitaille, T., {et~al.} 2022, Zenodo, \dodoi{10.5281/zenodo.6825092}

\bibitem[{{Brammer}(2023)}]{2023zndo...7712834B}
{Brammer}, G. 2023, {grizli}, 1.8.2, Zenodo,  Zenodo, \dodoi{10.5281/zenodo.7712834}

\bibitem[{Brammer {et~al.}(2008)Brammer, Dokkum, \& Coppi}]{brammer_eazy_2008}
Brammer, G.~B., Dokkum, P. G.~v., \& Coppi, P. 2008, ApJ, 686, 1503, \dodoi{10.1086/591786}

\bibitem[{{Briggs}(1995)}]{1995AAS...18711202B}
{Briggs}, D.~S. 1995, in American Astronomical Society Meeting Abstracts, Vol. 187, 112.02

\bibitem[{Calzetti {et~al.}(2000)Calzetti, Armus, Bohlin, Kinney, Koornneef, \& Storchi-Bergmann}]{calzetti_dust_2000}
Calzetti, D., Armus, L., Bohlin, R.~C., {et~al.} 2000, ApJ, 533, 682, \dodoi{10.1086/308692}

\bibitem[{Cappelluti {et~al.}(2009)Cappelluti, Brusa, Hasinger, Comastri, Zamorani, Finoguenov, Gilli, Puccetti, Miyaji, Salvato, Vignali, Aldcroft, Böhringer, Brunner, Civano, Elvis, Fiore, Fruscione, Griffiths, Guzzo, Iovino, Koekemoer, Mainieri, Scoville, Shopbell, Silverman, \& Urry}]{cappelluti_xmm-newton_2009}
Cappelluti, N., Brusa, M., Hasinger, G., {et~al.} 2009, A\&A, 497, 635, \dodoi{10.1051/0004-6361/200810794}

\bibitem[{Carilli {et~al.}(2010)Carilli, Daddi, Riechers, Walter, Weiss, Dannerbauer, Morrison, Wagg, Dav\'e, Elbaz, Stern, Dickinson, Krips, \& Aravena}]{carilli_imaging_2010}
Carilli, C.~L., Daddi, E., Riechers, D., {et~al.} 2010, ApJ, 714, 1407, \dodoi{10.1088/0004-637X/714/2/1407}

\bibitem[{Carnall {et~al.}(2018)Carnall, McLure, Dunlop, \& Dav\'e}]{carnall_inferring_2018}
Carnall, A.~C., McLure, R.~J., Dunlop, J.~S., \& Dav\'e, R. 2018, Monthly Notices of the Royal Astronomical Society, 480, 4379, \dodoi{10.1093/mnras/sty2169}

\bibitem[{Casey {et~al.}(2023)Casey, Kartaltepe, Drakos, Franco, Harish, Paquereau, Ilbert, Rose, Cox, Nightingale, Robertson, Silverman, Koekemoer, Massey, McCracken, Rhodes, Akins, Allen, Amvrosiadis, Arango-Toro, Bagley, Bongiorno, Capak, Champagne, Chartab, Ortiz, Chworowsky, Cooke, Cooper, Darvish, Ding, Faisst, Finkelstein, Fujimoto, Gentile, Gillman, Gould, Gozaliasl, Hayward, He, Hemmati, Hirschmann, Jahnke, Jin, Khostovan, Kokorev, Lambrides, Laigle, Larson, Leung, Liu, Liaudat, Long, Magdis, Mahler, Mainieri, Manning, Maraston, Martin, McCleary, McKinney, McPartland, Mobasher, Pattnaik, Renzini, Rich, Sanders, Sattari, Scognamiglio, Scoville, Sheth, Shuntov, Sparre, Suzuki, Talia, Toft, Trakhtenbrot, Urry, Valentino, Vanderhoof, Vardoulaki, Weaver, Whitaker, Wilkins, Yang, \& Zavala}]{casey_cosmos-web_2023}
Casey, C.~M., Kartaltepe, J.~S., Drakos, N.~E., {et~al.} 2023, ApJ, 954, 31, \dodoi{10.3847/1538-4357/acc2bc}

\bibitem[{Chabrier(2003)}]{chabrier_galactic_2003}
Chabrier, G. 2003, PASP, 115, 763, \dodoi{10.1086/376392}

\bibitem[{Civano {et~al.}(2016)Civano, Marchesi, Comastri, Urry, Elvis, Cappelluti, Puccetti, Brusa, Zamorani, Hasinger, Aldcroft, Alexander, Allevato, Brunner, Capak, Finoguenov, Fiore, Fruscione, Gilli, Glotfelty, Griffiths, Hao, Harrison, Jahnke, Kartaltepe, Karim, LaMassa, Lanzuisi, Miyaji, Ranalli, Salvato, Sargent, Scoville, Schawinski, Schinnerer, Silverman, Smolcic, Stern, Toft, Trakhenbrot, Treister, \& Vignali}]{civano_chandra_2016}
Civano, F., Marchesi, S., Comastri, A., {et~al.} 2016, ApJ, 819, 62, \dodoi{10.3847/0004-637X/819/1/62}

\bibitem[{Clark {et~al.}(2016)Clark, Schofield, Gomez, \& Davies}]{clark_empirical_2016}
Clark, C. J.~R., Schofield, S.~P., Gomez, H.~L., \& Davies, J.~I. 2016, Monthly Notices of the Royal Astronomical Society, 459, 1646, \dodoi{10.1093/mnras/stw647}

\bibitem[{{Conway} {et~al.}(1990){Conway}, {Cornwell}, \& {Wilkinson}}]{1990MNRAS.246..490C}
{Conway}, J.~E., {Cornwell}, T.~J., \& {Wilkinson}, P.~N. 1990, \mnras, 246, 490

\bibitem[{Cornwell(2008)}]{cornwell_multiscale_2008}
Cornwell, T.~J. 2008, IEEE J. Sel. Top. Signal Process., 2, 793, \dodoi{10.1109/JSTSP.2008.2006388}

\bibitem[{Daddi {et~al.}(2015)Daddi, Dannerbauer, Liu, Aravena, Bournaud, Walter, Riechers, Magdis, Sargent, B\'ethermin, Carilli, Cibinel, Dickinson, Elbaz, Gao, Gobat, Hodge, \& Krips}]{daddi_co_2015}
Daddi, E., Dannerbauer, H., Liu, D., {et~al.} 2015, A\&A, 577, A46, \dodoi{10.1051/0004-6361/201425043}

\bibitem[{Dekel \& Burkert(2014)}]{dekel_wet_2014}
Dekel, A., \& Burkert, A. 2014, Monthly Notices of the Royal Astronomical Society, 438, 1870, \dodoi{10.1093/mnras/stt2331}

\bibitem[{Dekel {et~al.}(2009{\natexlab{a}})Dekel, Sari, \& Ceverino}]{dekel_formation_2009}
Dekel, A., Sari, R., \& Ceverino, D. 2009{\natexlab{a}}, ApJ, 703, 785, \dodoi{10.1088/0004-637X/703/1/785}

\bibitem[{Dekel {et~al.}(2023)Dekel, Tziperman, Sarkar, Ginzburg, Mandelker, Ceverino, \& Primack}]{dekel_conditions_2023}
Dekel, A., Tziperman, O., Sarkar, K.~C., {et~al.} 2023, Monthly Notices of the Royal Astronomical Society, 521, 4299, \dodoi{10.1093/mnras/stad855}

\bibitem[{Dekel {et~al.}(2009{\natexlab{b}})Dekel, Birnboim, Engel, Freundlich, Goerdt, Mumcuoglu, Neistein, Pichon, Teyssier, \& Zinger}]{dekel_cold_2009}
Dekel, A., Birnboim, Y., Engel, G., {et~al.} 2009{\natexlab{b}}, Nature, 457, 451, \dodoi{10.1038/nature07648}

\bibitem[{{Ding} {et~al.}(2020){Ding}, {Silverman}, {Treu}, {Schulze}, {Schramm}, {Birrer}, {Park}, {Jahnke}, {Bennert}, {Kartaltepe}, {Koekemoer}, {Malkan}, \& {Sanders}}]{2020ApJ...888...37D}
{Ding}, X., {Silverman}, J., {Treu}, T., {et~al.} 2020, \apj, 888, 37, \dodoi{10.3847/1538-4357/ab5b90}

\bibitem[{Elbaz {et~al.}(2018)Elbaz, Leiton, Nagar, Okumura, Franco, Schreiber, Pannella, Wang, Dickinson, Díaz-Santos, Ciesla, Daddi, Bournaud, Magdis, Zhou, \& Rujopakarn}]{elbaz_starbursts_2018}
Elbaz, D., Leiton, R., Nagar, N., {et~al.} 2018, A\&A, 616, A110, \dodoi{10.1051/0004-6361/201732370}

\bibitem[{F\"orster~Schreiber \& Wuyts(2020)}]{forster_schreiber_star-forming_2020}
F\"orster~Schreiber, N.~M., \& Wuyts, S. 2020, ARA\&A, 58, 661, \dodoi{10.1146/annurev-astro-032620-021910}

\bibitem[{F\"orster~Schreiber {et~al.}(2018)F\"orster~Schreiber, Renzini, Mancini, Genzel, Bouch\'e, Cresci, Hicks, Lilly, Peng, Burkert, Carollo, Cimatti, Daddi, Davies, Genel, Kurk, Lang, Lutz, Mainieri, McCracken, Mignoli, Naab, Oesch, Pozzetti, Scodeggio, Griffin, Shapley, Sternberg, Tacchella, Tacconi, Wuyts, \& Zamorani}]{schreiber_sinszc-sinf_2018}
F\"orster~Schreiber, N.~M., Renzini, A., Mancini, C., {et~al.} 2018, ApJS, 238, 21, \dodoi{10.3847/1538-4365/aadd49}

\bibitem[{Freundlich {et~al.}(2013)Freundlich, Combes, Tacconi, Cooper, Genzel, Neri, Bolatto, Bournaud, Burkert, Cox, Davis, Schreiber, Garcia-Burillo, Gracia-Carpio, Lutz, Naab, Newman, Sternberg, \& Weiner}]{freundlich_towards_2013}
Freundlich, J., Combes, F., Tacconi, L.~J., {et~al.} 2013, A\&A, 553, A130, \dodoi{10.1051/0004-6361/201220981}

\bibitem[{Genzel {et~al.}(2010)Genzel, Tacconi, Gracia-Carpio, Sternberg, Cooper, Shapiro, Bolatto, Bouch\'e, Bournaud, Burkert, Combes, Comerford, Cox, Davis, Schreiber, Garcia-Burillo, Lutz, Naab, Neri, Omont, Shapley, \& Weiner}]{genzel_study_2010}
Genzel, R., Tacconi, L.~J., Gracia-Carpio, J., {et~al.} 2010, Monthly Notices of the Royal Astronomical Society, 407, 2091, \dodoi{10.1111/j.1365-2966.2010.16969.x}

\bibitem[{G{\'o}mez-Guijarro {et~al.}(2022)G{\'o}mez-Guijarro, Elbaz, Xiao, Kokorev, Magdis, Magnelli, Daddi, Valentino, Sargent, Dickinson, Béthermin, Franco, Pope, Kalita, Ciesla, Demarco, Inami, Rujopakarn, Shu, Wang, Zhou, Alexander, Bournaud, Chary, Ferguson, Finkelstein, Giavalisco, Iono, Juneau, Kartaltepe, Lagache, Floc’h, Leiton, Leroy, Lin, Motohara, Mullaney, Okumura, Pannella, Papovich, \& Treister}]{gomez-guijarro_goods-alma_2022}
G{\'o}mez-Guijarro, C., Elbaz, D., Xiao, M., {et~al.} 2022, A\&A, 659, A196, \dodoi{10.1051/0004-6361/202142352}

\bibitem[{Gonz\'alez-L\'opez {et~al.}(2017)Gonz\'alez-L\'opez, Bauer, Aravena, Laporte, Bradley, Carrasco, Carvajal, Demarco, Infante, Kneissl, Koekemoer, Arancibia, Troncoso, Villard, \& Zitrin}]{gonzalez-lopez_alma_2017}
Gonz\'alez-L\'opez, J., Bauer, F.~E., Aravena, M., {et~al.} 2017, A\&A, 608, A138, \dodoi{10.1051/0004-6361/201730961}

\bibitem[{Greve {et~al.}(2014)Greve, Leonidaki, Xilouris, Weiß, Zhang, van~der Werf, Aalto, Armus, Díaz-Santos, Evans, Fischer, Gao, Gonz\'alez-Alfonso, Harris, Henkel, Meijerink, Naylor, Smith, Spaans, Stacey, Veilleux, \& Walter}]{greve_star_2014}
Greve, T.~R., Leonidaki, I., Xilouris, E.~M., {et~al.} 2014, The Astrophysical Journal, 794, 142, \dodoi{10.1088/0004-637X/794/2/142}

\bibitem[{Hao {et~al.}(2011)Hao, Kennicutt, Johnson, Calzetti, Dale, \& Moustakas}]{hao_dust-corrected_2011}
Hao, C.-N., Kennicutt, R.~C., Johnson, B.~D., {et~al.} 2011, ApJ, 741, 124, \dodoi{10.1088/0004-637X/741/2/124}

\bibitem[{Hernquist(1989)}]{hernquist_tidal_1989}
Hernquist, L. 1989, Nature, 340, 687, \dodoi{10.1038/340687a0}

\bibitem[{Huang {et~al.}(2023)Huang, Li, Cheng, Hou, Yan, Willner, Dai, Zheng, Pan, Rigopoulou, Wang, Li, Liang, Esamdin, \& Fazio}]{huang_diverse_2023}
Huang, J.-S., Li, Z.-J., Cheng, C., {et~al.} 2023, ApJ, 949, 83, \dodoi{10.3847/1538-4357/acc9c3}

\bibitem[{Ilbert {et~al.}(2006)Ilbert, Arnouts, McCracken, Bolzonella, Bertin, Fèvre, Mellier, Zamorani, Pellò, Iovino, Tresse, Brun, Bottini, Garilli, Maccagni, Picat, Scaramella, Scodeggio, Vettolani, Zanichelli, Adami, Bardelli, Cappi, Charlot, Ciliegi, Contini, Cucciati, Foucaud, Franzetti, Gavignaud, Guzzo, Marano, Marinoni, Mazure, Meneux, Merighi, Paltani, Pollo, Pozzetti, Radovich, Zucca, Bondi, Bongiorno, Busarello, Torre, Gregorini, Lamareille, Mathez, Merluzzi, Ripepi, Rizzo, \& Vergani}]{ilbert_accurate_2006}
Ilbert, O., Arnouts, S., McCracken, H.~J., {et~al.} 2006, A\&A, 457, 841, \dodoi{10.1051/0004-6361:20065138}

\bibitem[{Johnson {et~al.}(2018)Johnson, Harrison, Swinbank, Tiley, Stott, Bower, Smail, Bunker, Sobral, Turner, Best, Bureau, Cirasuolo, Jarvis, Magdis, Sharples, Bland-Hawthorn, Catinella, Cortese, Croom, Federrath, Glazebrook, Sweet, Bryant, Goodwin, Konstantopoulos, Lawrence, Medling, Owers, \& Richards}]{johnson_kmos_2018}
Johnson, H.~L., Harrison, C.~M., Swinbank, A.~M., {et~al.} 2018, Monthly Notices of the Royal Astronomical Society, 474, 5076, \dodoi{10.1093/mnras/stx3016}

\bibitem[{Kalita {et~al.}(2023)Kalita, Silverman, Daddi, Bottrell, Ho, Ding, \& Yang}]{kalita_rest-frame_2023}
Kalita, B.~S., Silverman, J.~D., Daddi, E., {et~al.} 2023, ApJ, 960, 25, \dodoi{10.3847/1538-4357/acfee4}

\bibitem[{Kalita {et~al.}(2024)Kalita, Silverman, Daddi, Mercier, Ho, \& Ding}]{kalita_near-ir_2024}
---. 2024, Near-{IR} clumps and their properties in high-z galaxies with {JWST}/{NIRCam},  arXiv, \dodoi{10.48550/arXiv.2402.02679}

\bibitem[{{Kashino} {et~al.}(2019){Kashino}, {Silverman}, {Sanders}, {Kartaltepe}, {Daddi}, {Renzini}, {Rodighiero}, {Puglisi}, {Valentino}, {Juneau}, {Arimoto}, {Nagao}, {Ilbert}, {Le F{\`e}vre}, \& {Koekemoer}}]{Kashino2019}
{Kashino}, D., {Silverman}, J.~D., {Sanders}, D., {et~al.} 2019, \apjs, 241, 10, \dodoi{10.3847/1538-4365/ab06c4}

\bibitem[{Kennicutt(1998{\natexlab{a}})}]{robert_c_kennicutt_global_1998}
Kennicutt. 1998{\natexlab{a}}, ApJ, 498, 541, \dodoi{10.1086/305588}

\bibitem[{Kennicutt(1998{\natexlab{b}})}]{kennicutt_star_1998}
Kennicutt, R.~C. 1998{\natexlab{b}}, ARA\&A, 36, 189, \dodoi{10.1146/annurev.astro.36.1.189}

\bibitem[{Koekemoer {et~al.}(2007)Koekemoer, Aussel, Calzetti, Capak, Giavalisco, Kneib, Leauthaud, Fèvre, McCracken, Massey, Mobasher, Rhodes, Scoville, \& Shopbell}]{koekemoer_cosmos_2007}
Koekemoer, A.~M., Aussel, H., Calzetti, D., {et~al.} 2007, ApJS, 172, 196, \dodoi{10.1086/520086}

\bibitem[{Koekemoer {et~al.}(2011)Koekemoer, Faber, Ferguson, Grogin, Kocevski, Koo, Lai, Lotz, Lucas, McGrath, Ogaz, Rajan, Riess, Rodney, Strolger, Casertano, Castellano, Dahlen, Dickinson, Dolch, Fontana, Giavalisco, Grazian, Guo, Hathi, Huang, Wel, Yan, Acquaviva, Alexander, Almaini, Ashby, Barden, Bell, Bournaud, Brown, Caputi, Cassata, Challis, Chary, Cheung, Cirasuolo, Conselice, Cooray, Croton, Daddi, Dav\'e, Mello, Ravel, Dekel, Donley, Dunlop, Dutton, Elbaz, Fazio, Filippenko, Finkelstein, Frazer, Gardner, Garnavich, Gawiser, Gruetzbauch, Hartley, Häussler, Herrington, Hopkins, Huang, Jha, Johnson, Kartaltepe, Khostovan, Kirshner, Lani, Lee, Li, Madau, McCarthy, McIntosh, McLure, McPartland, Mobasher, Moreira, Mortlock, Moustakas, Mozena, Nandra, Newman, Nielsen, Niemi, Noeske, Papovich, Pentericci, Pope, Primack, Ravindranath, Reddy, Renzini, Rix, Robaina, Rosario, Rosati, Salimbeni, Scarlata, Siana, Simard, Smidt, Snyder, Somerville, Spinrad, Straughn, Telford, Teplitz, Trump, Vargas, Villforth,
  Wagner, Wandro, Wechsler, Weiner, Wiklind, Wild, Wilson, Wuyts, \& Yun}]{koekemoer_candels_2011}
Koekemoer, A.~M., Faber, S.~M., Ferguson, H.~C., {et~al.} 2011, ApJS, 197, 36, \dodoi{10.1088/0067-0049/197/2/36}

\bibitem[{Lackner {et~al.}(2014)Lackner, Silverman, Salvato, Kampczyk, Kartaltepe, Sanders, Capak, Civano, Halliday, Ilbert, Jahnke, Koekemoer, Lee, Le~Fèvre, Liu, Scoville, Sheth, \& Toft}]{lackner_late-stage_2014}
Lackner, C.~N., Silverman, J.~D., Salvato, M., {et~al.} 2014, ApJ, 148, 137, \dodoi{10.1088/0004-6256/148/6/137}

\bibitem[{{Le Bail} {et~al.}(2023){Le Bail}, {Daddi}, {Elbaz}, {Dickinson}, {Giavalisco}, {Magnelli}, {G{\'o}mez-Guijarro}, {Kalita}, {Koekemoer}, {Holwerda}, {Bournaud}, {de la Vega}, {Calabr{\`o}}, {Dekel}, {Cheng}, {Bisigello}, {Franco}, {Costantin}, {Lucas}, {P{\'e}rez-Gonz{\'a}lez}, {Lu}, {Wilkins}, {Arrabal Haro}, {Bagley}, {Finkelstein}, {Kartaltepe}, {Papovich}, {Pirzkal}, \& {Yung}}]{2023arXiv230707599L}
{Le Bail}, A., {Daddi}, E., {Elbaz}, D., {et~al.} 2023, arXiv e-prints, arXiv:2307.07599, \dodoi{10.48550/arXiv.2307.07599}

\bibitem[{Lin {et~al.}(2022)Lin, Ellison, Pan, Thorp, Yu, Belfiore, Hsieh, Maiolino, Ramya, S\'anchez, \& Su}]{lin_almaquest_2022}
Lin, L., Ellison, S.~L., Pan, H.-A., {et~al.} 2022, ApJ, 926, 175, \dodoi{10.3847/1538-4357/ac4ccc}

\bibitem[{Liu {et~al.}(2015)Liu, Gao, Isaak, Daddi, Yang, Lu, \& Werf}]{liu_high-j_2015}
Liu, D., Gao, Y., Isaak, K., {et~al.} 2015, ApJL, 810, L14, \dodoi{10.1088/2041-8205/810/2/L14}

\bibitem[{Liu {et~al.}(2019)Liu, Lang, Magnelli, Schinnerer, Leslie, Fudamoto, Bondi, Groves, Jim\'enez-Andrade, Harrington, Karim, Oesch, Sargent, Vardoulaki, Bǎdescu, Moser, Bertoldi, Battisti, Cunha, Zavala, Vaccari, Davidzon, Riechers, \& Aravena}]{liu_automated_2019}
Liu, D., Lang, P., Magnelli, B., {et~al.} 2019, ApJS, 244, 40, \dodoi{10.3847/1538-4365/ab42da}

\bibitem[{Liu {et~al.}(2021)Liu, Daddi, Schinnerer, Saito, Leroy, Silverman, Valentino, Magdis, Gao, Jin, Puglisi, \& Groves}]{liu_co_2021}
Liu, D., Daddi, E., Schinnerer, E., {et~al.} 2021, ApJ, 909, 56, \dodoi{10.3847/1538-4357/abd801}

\bibitem[{{Lotz} {et~al.}(2008){Lotz}, {Davis}, {Faber}, {Guhathakurta}, {Gwyn}, {Huang}, {Koo}, {Le Floc'h}, {Lin}, {Newman}, {Noeske}, {Papovich}, {Willmer}, {Coil}, {Conselice}, {Cooper}, {Hopkins}, {Metevier}, {Primack}, {Rieke}, \& {Weiner}}]{Lotz2008}
{Lotz}, J.~M., {Davis}, M., {Faber}, S.~M., {et~al.} 2008, \apj, 672, 177, \dodoi{10.1086/523659}

\bibitem[{Magdis {et~al.}(2012)Magdis, Daddi, B\'ethermin, Sargent, Elbaz, Pannella, Dickinson, Dannerbauer, Cunha, Walter, Rigopoulou, Charmandaris, Hwang, \& Kartaltepe}]{magdis_evolving_2012}
Magdis, G.~E., Daddi, E., B\'ethermin, M., {et~al.} 2012, ApJ, 760, 6, \dodoi{10.1088/0004-637X/760/1/6}

\bibitem[{Marchesi {et~al.}(2016)Marchesi, Civano, Elvis, Salvato, Brusa, Comastri, Gilli, Hasinger, Lanzuisi, Miyaji, Treister, Urry, Vignali, Zamorani, Allevato, Cappelluti, Cardamone, Finoguenov, Griffiths, Karim, Laigle, LaMassa, Jahnke, Ranalli, Schawinski, Schinnerer, Silverman, Smolcic, Suh, \& Trakhtenbrot}]{marchesi_chandra_2016}
Marchesi, S., Civano, F., Elvis, M., {et~al.} 2016, ApJ, 817, 34, \dodoi{10.3847/0004-637X/817/1/34}

\bibitem[{Masters {et~al.}(2019)Masters, Stern, Cohen, Capak, Stanford, Hernitschek, Galametz, Davidzon, Rhodes, Sanders, Mobasher, Castander, Pruett, \& Fotopoulou}]{masters_complete_2019}
Masters, D.~C., Stern, D.~K., Cohen, J.~G., {et~al.} 2019, ApJ, 877, 81, \dodoi{10.3847/1538-4357/ab184d}

\bibitem[{Mihos \& Hernquist(1994)}]{mihos_ultraluminous_1994}
Mihos, J.~C., \& Hernquist, L. 1994, ApJL, 431, L9, \dodoi{10.1086/187460}

\bibitem[{Moreno {et~al.}(2021)Moreno, Torrey, Ellison, Patton, Bottrell, Bluck, Hani, Hayward, Bullock, Hopkins, \& Hernquist}]{moreno_spatially_2021}
Moreno, J., Torrey, P., Ellison, S.~L., {et~al.} 2021, MNRAS, 503, 3113, \dodoi{10.1093/mnras/staa2952}

\bibitem[{Mullaney {et~al.}(2011)Mullaney, Alexander, Goulding, \& Hickox}]{mullaney_defining_2011}
Mullaney, J.~R., Alexander, D.~M., Goulding, A.~D., \& Hickox, R.~C. 2011, Monthly Notices of the Royal Astronomical Society, 414, 1082, \dodoi{10.1111/j.1365-2966.2011.18448.x}

\bibitem[{Murphy {et~al.}(2011)Murphy, Condon, Schinnerer, Kennicutt, Calzetti, Armus, Helou, Turner, Aniano, Beirão, Bolatto, Brandl, Croxall, Dale, Meyer, Draine, Engelbracht, Hunt, Hao, Koda, Roussel, Skibba, \& Smith}]{murphy_calibrating_2011}
Murphy, E.~J., Condon, J.~J., Schinnerer, E., {et~al.} 2011, ApJ, 737, 67, \dodoi{10.1088/0004-637X/737/2/67}

\bibitem[{Osborne \& Salim(2024)}]{osborne_strategies_2024}
Osborne, C., \& Salim, S. 2024, Strategies for obtaining robust {SED} fitting parameters for galaxies at z{\textasciitilde}1 and z{\textasciitilde}2 in the absence of {IR} data,  arXiv, \dodoi{10.48550/arXiv.2401.06865}

\bibitem[{Puglisi {et~al.}(2019)Puglisi, Daddi, Liu, Bournaud, Silverman, Circosta, Calabrò, Aravena, Cibinel, Dannerbauer, Delvecchio, Elbaz, Gao, Gobat, Jin, Floc'h, Magdis, Mancini, Riechers, Rodighiero, Sargent, Valentino, \& Zanisi}]{puglisi_main_2019}
Puglisi, A., Daddi, E., Liu, D., {et~al.} 2019, ApJL, 877, L23, \dodoi{10.3847/2041-8213/ab1f92}

\bibitem[{Puglisi {et~al.}(2021)Puglisi, Daddi, Valentino, Magdis, Liu, Kokorev, Circosta, Elbaz, Bournaud, Gomez-Guijarro, Jin, Madden, Sargent, \& Swinbank}]{puglisi_sub-millimetre_2021}
Puglisi, A., Daddi, E., Valentino, F., {et~al.} 2021, MNRAS, 508, 5217, \dodoi{10.1093/mnras/stab2914}

\bibitem[{Rizzo {et~al.}(2023)Rizzo, Roman-Oliveira, Fraternali, Frickmann, Valentino, Brammer, Zanella, Kokorev, Popping, Whitaker, Kohandel, Magdis, Di~Mascolo, Ikeda, Jin, \& Toft}]{rizzo_alma-alpaka_2023}
Rizzo, F., Roman-Oliveira, F., Fraternali, F., {et~al.} 2023, The {ALMA}-{ALPAKA} survey {I}: high-resolution {CO} and [{CI}] kinematics of star-forming galaxies at z = 0.5-3.5,  arXiv.
\newblock \url{http://arxiv.org/abs/2303.16227}

\bibitem[{Rodighiero {et~al.}(2011)Rodighiero, Daddi, Baronchelli, Cimatti, Renzini, Aussel, Popesso, Lutz, Andreani, Berta, Cava, Elbaz, Feltre, Fontana, Schreiber, Franceschini, Genzel, Grazian, Gruppioni, Ilbert, Floch, Magdis, Magliocchetti, Magnelli, Maiolino, McCracken, Nordon, Poglitsch, Santini, Pozzi, Riguccini, Tacconi, Wuyts, \& Zamorani}]{rodighiero_lesser_2011}
Rodighiero, G., Daddi, E., Baronchelli, I., {et~al.} 2011, ApJL, 739, L40, \dodoi{10.1088/2041-8205/739/2/L40}

\bibitem[{Rujopakarn {et~al.}(2023)Rujopakarn, Williams, Daddi, Schramm, Sun, Alberts, Rieke, Tan, Tacchella, Giavalisco, \& Silverman}]{rujopakarn_jwst_2023}
Rujopakarn, W., Williams, C.~C., Daddi, E., {et~al.} 2023, ApJL, 948, L8, \dodoi{10.3847/2041-8213/accc82}

\bibitem[{Sanders {et~al.}(1988)Sanders, Soifer, Elias, Madore, Matthews, Neugebauer, \& Scoville}]{sanders_ultraluminous_1988}
Sanders, D.~B., Soifer, B.~T., Elias, J.~H., {et~al.} 1988, ApJ, 325, 74, \dodoi{10.1086/165983}

\bibitem[{Schmidt(1959)}]{schmidt_rate_1959}
Schmidt, M. 1959, ApJ, 129, 243, \dodoi{10.1086/146614}

\bibitem[{Scoville {et~al.}(2007)Scoville, Abraham, Aussel, Barnes, Benson, Blain, Calzetti, Comastri, Capak, Carilli, Carlstrom, Carollo, Colbert, Daddi, Ellis, Elvis, Ewald, Fall, Franceschini, Giavalisco, Green, Griffiths, Guzzo, Hasinger, Impey, Kneib, Koda, Koekemoer, Lefevre, Lilly, Liu, McCracken, Massey, Mellier, Miyazaki, Mobasher, Mould, Norman, Refregier, Renzini, Rhodes, Rich, Sanders, Schiminovich, Schinnerer, Scodeggio, Sheth, Shopbell, Taniguchi, Tyson, Urry, Waerbeke, Vettolani, White, \& Yan}]{scoville_cosmos_2007}
Scoville, N., Abraham, R.~G., Aussel, H., {et~al.} 2007, ApJS, 172, 38, \dodoi{10.1086/516580}

\bibitem[{Scoville {et~al.}(2016)Scoville, Sheth, Aussel, Bout, Capak, Bongiorno, Casey, Murchikova, Koda, \'alvarez M\'arquez, Lee, Laigle, McCracken, Ilbert, Pope, Sanders, Chu, Toft, Ivison, \& Manohar}]{scoville_ism_2016}
Scoville, N., Sheth, K., Aussel, H., {et~al.} 2016, ApJ, 820, 83, \dodoi{10.3847/0004-637X/820/2/83}

\bibitem[{Scoville {et~al.}(2023)Scoville, Faisst, Weaver, Toft, McCracken, Ilbert, Diaz-Santos, Staguhn, Koda, Casey, Sanders, Mobasher, Chartab, Sattari, Capak, Bout, Bongiorno, Vlahakis, Sheth, Yun, Aussel, Laigle, \& Masters}]{scoville_cosmic_2023}
Scoville, N., Faisst, A., Weaver, J., {et~al.} 2023, ApJ, 943, 82, \dodoi{10.3847/1538-4357/aca1bc}

\bibitem[{Silverman {et~al.}(2015{\natexlab{a}})Silverman, Daddi, Rodighiero, Rujopakarn, Sargent, Renzini, Liu, Feruglio, Kashino, Sanders, Kartaltepe, Nagao, Arimoto, Berta, B\'ethermin, Koekemoer, Lutz, Magdis, Mancini, Onodera, \& Zamorani}]{silverman_higher_2015}
Silverman, J.~D., Daddi, E., Rodighiero, G., {et~al.} 2015{\natexlab{a}}, ApJL, 812, L23, \dodoi{10.1088/2041-8205/812/2/L23}

\bibitem[{Silverman {et~al.}(2015{\natexlab{b}})Silverman, Kashino, Sanders, Kartaltepe, Arimoto, Renzini, Rodighiero, Daddi, Zahid, Nagao, Kewley, Lilly, Sugiyama, Baronchelli, Capak, Carollo, Chu, Hasinger, Ilbert, Juneau, Kajisawa, Koekemoer, Kovac, Fèvre, Masters, McCracken, Onodera, Schulze, Scoville, Strazzullo, \& Taniguchi}]{silverman_fmos-cosmos_2015}
Silverman, J.~D., Kashino, D., Sanders, D., {et~al.} 2015{\natexlab{b}}, ApJS, 220, 12, \dodoi{10.1088/0067-0049/220/1/12}

\bibitem[{Silverman {et~al.}(2018{\natexlab{a}})Silverman, Rujopakarn, Daddi, Renzini, Rodighiero, Liu, Puglisi, Sargent, Mancini, Kartaltepe, Kashino, Koekemoer, Arimoto, B\'ethermin, Jin, Magdis, Nagao, Onodera, Sanders, \& Valentino}]{silverman_molecular_2018}
Silverman, J.~D., Rujopakarn, W., Daddi, E., {et~al.} 2018{\natexlab{a}}, ApJ, 867, 92, \dodoi{10.3847/1538-4357/aae25e}

\bibitem[{Silverman {et~al.}(2018{\natexlab{b}})Silverman, Daddi, Rujopakarn, Renzini, Mancini, Bournaud, Puglisi, Rodighiero, Liu, Sargent, Arimoto, B\'ethermin, Fensch, Hayward, Kartaltepe, Kashino, Koekemoer, Magdis, McCracken, Nagao, Sheth, Smolčić, \& Valentino}]{silverman_concurrent_2018}
Silverman, J.~D., Daddi, E., Rujopakarn, W., {et~al.} 2018{\natexlab{b}}, ApJ, 868, 75, \dodoi{10.3847/1538-4357/aae64b}

\bibitem[{Solomon {et~al.}(1997)Solomon, Downes, Radford, \& Barrett}]{solomon_molecular_1997}
Solomon, P.~M., Downes, D., Radford, S. J.~E., \& Barrett, J.~W. 1997, ApJ, 478, 144, \dodoi{10.1086/303765}

\bibitem[{Speagle {et~al.}(2014)Speagle, Steinhardt, Capak, \& Silverman}]{speagle_highly_2014}
Speagle, J.~S., Steinhardt, C.~L., Capak, P.~L., \& Silverman, J.~D. 2014, ApJS, 214, 15, \dodoi{10.1088/0067-0049/214/2/15}

\bibitem[{Tacchella {et~al.}(2016{\natexlab{a}})Tacchella, Dekel, Carollo, Ceverino, DeGraf, Lapiner, Mandelker, \& Primack}]{tacchella_evolution_2016}
Tacchella, S., Dekel, A., Carollo, C.~M., {et~al.} 2016{\natexlab{a}}, Monthly Notices of the Royal Astronomical Society, 458, 242, \dodoi{10.1093/mnras/stw303}

\bibitem[{Tacchella {et~al.}(2016{\natexlab{b}})Tacchella, Dekel, Carollo, Ceverino, DeGraf, Lapiner, Mandelker, \& Primack~Joel}]{tacchella_confinement_2016}
---. 2016{\natexlab{b}}, MNRAS, 457, 2790, \dodoi{10.1093/mnras/stw131}

\bibitem[{Tacconi {et~al.}(2020)Tacconi, Genzel, \& Sternberg}]{tacconi_evolution_2020}
Tacconi, L.~J., Genzel, R., \& Sternberg, A. 2020, Annual Review of Astronomy and Astrophysics, 58, 157, \dodoi{10.1146/annurev-astro-082812-141034}

\bibitem[{Tacconi {et~al.}(2013)Tacconi, Neri, Genzel, Combes, Bolatto, Cooper, Wuyts, Bournaud, Burkert, Comerford, Cox, Davis, Schreiber, García-Burillo, Gracia-Carpio, Lutz, Naab, Newman, Omont, Saintonge, Griffin, Shapley, Sternberg, \& Weiner}]{tacconi_phibss_2013}
Tacconi, L.~J., Neri, R., Genzel, R., {et~al.} 2013, ApJ, 768, 74, \dodoi{10.1088/0004-637X/768/1/74}

\bibitem[{Tacconi {et~al.}(2018)Tacconi, Genzel, Saintonge, Combes, García-Burillo, Neri, Bolatto, Contini, Schreiber, Lilly, Lutz, Wuyts, Accurso, Boissier, Boone, Bouch\'e, Bournaud, Burkert, Carollo, Cooper, Cox, Feruglio, Freundlich, Herrera-Camus, Juneau, Lippa, Naab, Renzini, Salome, Sternberg, Tadaki, \"Ubler, Walter, Weiner, \& Weiss}]{tacconi_phibss_2018}
Tacconi, L.~J., Genzel, R., Saintonge, A., {et~al.} 2018, ApJ, 853, 179, \dodoi{10.3847/1538-4357/aaa4b4}

\bibitem[{{Tadaki} {et~al.}(2020){Tadaki}, {Belli}, {Burkert}, {Dekel}, {F{\"o}rster Schreiber}, {Genzel}, {Hayashi}, {Herrera-Camus}, {Kodama}, {Kohno}, {Koyama}, {Lee}, {Lutz}, {Mowla}, {Nelson}, {Renzini}, {Suzuki}, {Tacconi}, {{\"U}bler}, {Wisnioski}, \& {Wuyts}}]{Tadaki2020}
{Tadaki}, K.-i., {Belli}, S., {Burkert}, A., {et~al.} 2020, \apj, 901, 74, \dodoi{10.3847/1538-4357/abaf4a}

\bibitem[{Teodoro \& Fraternali(2015)}]{teodoro_3dbarolo_2015}
Teodoro, E. M.~D., \& Fraternali, F. 2015, MNRAS, 451, 3021, \dodoi{10.1093/mnras/stv1213}

\bibitem[{\"Ubler {et~al.}(2018)\"Ubler, Genzel, Tacconi, Schreiber, Neri, Contursi, Belli, Nelson, Lang, Shimizu, Davies, Herrera-Camus, Lutz, Plewa, Price, Schuster, Sternberg, Tadaki, Wisnioski, \& Wuyts}]{ubler_ionized_2018}
\"Ubler, H., Genzel, R., Tacconi, L.~J., {et~al.} 2018, ApJL, 854, L24, \dodoi{10.3847/2041-8213/aaacfa}

\bibitem[{Valentino {et~al.}(2020)Valentino, Daddi, Puglisi, Magdis, Liu, Kokorev, Cortzen, Madden, Aravena, G\'omez-Guijarro, Lee, Floc’h, Gao, Gobat, Bournaud, Dannerbauer, Jin, Dickinson, Kartaltepe, \& Sanders}]{valentino_co_2020}
Valentino, F., Daddi, E., Puglisi, A., {et~al.} 2020, A\&A, 641, A155, \dodoi{10.1051/0004-6361/202038322}

\bibitem[{Weaver {et~al.}(2022)Weaver, Kauffmann, Ilbert, McCracken, Moneti, Toft, Brammer, Shuntov, Davidzon, Hsieh, Laigle, Anastasiou, Jespersen, Vinther, Capak, Casey, McPartland, Milvang-Jensen, Mobasher, Sanders, Zalesky, Arnouts, Aussel, Dunlop, Faisst, Franx, Furtak, Fynbo, Gould, Greve, Gwyn, Kartaltepe, Kashino, Koekemoer, Kokorev, Fèvre, Lilly, Masters, Magdis, Mehta, Peng, Riechers, Salvato, Sawicki, Scarlata, Scoville, Shirley, Silverman, Sneppen, Smolc̆ić, Steinhardt, Stern, Tanaka, Taniguchi, Teplitz, Vaccari, Wang, \& Zamorani}]{weaver_cosmos2020_2022}
Weaver, J.~R., Kauffmann, O.~B., Ilbert, O., {et~al.} 2022, ApJS, 258, 11, \dodoi{10.3847/1538-4365/ac3078}

\bibitem[{Whitaker {et~al.}(2012)Whitaker, Dokkum, Brammer, \& Franx}]{whitaker_star_2012}
Whitaker, K.~E., Dokkum, P. G.~v., Brammer, G., \& Franx, M. 2012, ApJL, 754, L29, \dodoi{10.1088/2041-8205/754/2/L29}

\bibitem[{Wisnioski {et~al.}(2019)Wisnioski, Schreiber, Fossati, Mendel, Wilman, Genzel, Bender, Wuyts, Davies, \"Ubler, Bandara, Beifiori, Belli, Brammer, Chan, Davies, Fabricius, Galametz, Lang, Lutz, Nelson, Momcheva, Price, Rosario, Saglia, Seitz, Shimizu, Tacconi, Tadaki, Dokkum, \& Wuyts}]{wisnioski_kmos3d_2019}
Wisnioski, E., Schreiber, N. M.~F., Fossati, M., {et~al.} 2019, ApJ, 886, 124, \dodoi{10.3847/1538-4357/ab4db8}

\end{thebibliography}
\bibliographystyle{aasjournal}

\end{document}